\documentclass[12pt,a4paper]{article}
\usepackage[utf8]{inputenc}
\usepackage[english]{babel}
\usepackage{graphicx}
\usepackage{amsmath}
\usepackage{amsthm}
\usepackage{amsfonts}
\usepackage{amssymb}
\usepackage{geometry}
\usepackage[longnamesfirst,authoryear]{natbib}
\usepackage{afterpage}
\usepackage{bm}
\usepackage{here}
\usepackage[subrefformat=parens]{subcaption}
\usepackage{cancel}
\usepackage{xcolor}
\usepackage{subcaption}

\newtheorem{Theorem}{Theorem}

\newtheorem{Lemma}{Lemma}
\newtheorem{Proposition}{Proposition}
\newtheorem{Assumption}{Assumption}

\newlength\tindent
\begin{document}

\author{Yagan Hazard\footnote{Collegio Carlo Alberto and University of Turin. Email: yagan.hazard@carloalberto.org}
\and 
Toru Kitagawa\footnote{Brown University, Department of Economics. Email: toru\_kitagawa@brown.edu} 
}

\title{Who With Whom? Learning Optimal Matching Policies\thanks{We would like to thank Thomas Carr, Yanqin Fan, Alfred Galichon, and Florian Gunsilius for beneficial comments. We also thank participants to the Collegio Carlo Alberto workshop on ``optimal transport in econometrics'' and the seminars at CUHK Shenzhen, Peking University, and UW Seattle. Toru Kitagawa thanks the Visiting Researcher Fellowship from Collegio Carlo Alberto.}}
\date{17 July 2025}
\maketitle

\begin{abstract}
There are many economic contexts where the productivity and welfare performance of institutions and policies depend on who matches with whom. 
Examples include caseworkers and job seekers in job search assistance programs, medical doctors and patients, teachers and students, attorneys and defendants, and tax auditors and taxpayers, among others. 
Although reallocating individuals through a change in matching policy can be less costly than training personnel or introducing a new program, methods for learning optimal matching policies and their statistical performance are less studied than methods for other policy interventions. 
This paper develops a method to learn welfare optimal matching policies for two-sided matching problems in which a planner matches individuals based on the rich set of observable characteristics of the two sides.
We formulate the learning problem as an empirical optimal transport problem with a match cost function estimated from training data, and propose estimating an optimal matching policy by maximizing the entropy regularized empirical welfare criterion. 
We derive a welfare regret bound for the estimated policy and characterize its convergence. 
We apply our proposal to the problem of matching caseworkers and job seekers in a job search assistance program, and assess its welfare performance in a simulation study calibrated with French administrative data.
\end{abstract}


\pagebreak

\section{Introduction}
Active job market programs commonly involve caseworkers providing advice and support to unemployed workers during their job search process. Empirical research such as \cite{dromundo2022} indicates that there is substantial heterogeneity in both job seekers and caseworkers, and the overall performance of these programs is driven by who matches with whom. 
This suggests an avenue for policy intervention. Instead of allocating caseworkers randomly, as is currently common, a planner can centrally assign caseworkers in a sophisticated manner, taking into account the complementarity or substitutability of the characteristics of caseworkers and job seekers. A similar opportunity for policy intervention exists in many other contexts including matching medical doctors and patients \citep{dahlstrand2024} or hospitals \citep{mourot2025}, teachers and classrooms \citep{graham2023}, attorneys and defendants \citep{spurr1987}, and tax auditors and taxpayers \citep{bergeron2025} among others. 

As emphasized by \cite{graham2014, graham2020}, reallocating individuals through a change in matching procedure can be less costly than other policy interventions, such as training or hiring personnel or introducing a new program. 
Methods for learning optimal matching policies therefore have the potential to have a large real-world welfare impact in a wide scope of applications. However, methods for efficiently learning an optimal matching policy from data and their statistical performance are not well studied. 

This paper develops a method to learn welfare optimal matching policies for two-sided matching problems in which a planner decides who should match with whom based on the observable characteristics of the two sides.
We formulate the learning problem as an empirical optimal transport problem with a match cost function estimated from training data. 
In this formulation, we ultimately uncover the joint distribution which minimizes the average match cost and has marginal distributions that coincide with the empirical distributions of observable characteristics for the two sides.
In the standard optimal transport formulation, an optimal matching policy can be obtained by linear programming. 
The dimension of choice variables in this linear program is $n^2$, where $n$ is the number of pairs created in the matching process. 

To keep the estimation of an optimal matching policy feasible even for large $n$, we propose maximizing the entropy regularized transport cost criterion with the Sinkhorn algorithm (iterated projection fitting). This computationally attractive procedure is proposed in \cite{Cuturi2013} and \cite{Galichon_Salanie_2022}, where its analytical properties are also studied. In the case of matching caseworkers and job seekers, the number of matches to be formed in a given month within each unemployment agency ranges from a few hundred to a few thousand. The corresponding unregularized optimal transportation problem is a linear program with 100,000 choice variables, which is computationally demanding. Solving the regularized problem with the Sinkhorn algorithm, on the other hand, involves iterations of arithmetic operations, which drastically reduces the complexity of computation.

A novel contribution of this article is to derive a welfare regret bound for the estimated matching policy and characterize its convergence, taking into account that the cost function is estimated. In contrast to the existing work on empirical optimal transport, e.g., \cite{Rigollet2022}, in which the cost function is assumed to be known, our welfare regret bound accounts for estimation errors and sampling uncertainty. Our bound is non-asymptotic, and we derive it under a minimal regularity condition. It exhibits a bias-variance tradeoff with respect to a regularization parameter, which can guide the choice of a value for the regularization parameter.  

We perform an extensive numerical study to assess the performance of our proposed method. A first simulation exercise with a simple data generating process illustrates how the welfare gains associated with our method depend on (i) the size of the training sample used to estimate (nonparametrically) the cost function and (ii) the choice of regularization parameter for the regularized optimal transport problem. In addition, to gauge the magnitude of the welfare gain of our proposal in a real-world scenario, we apply our method to the assignment of caseworkers to job seekers in a job search assistance program. We calibrate a data generating processes using estimates of key parameters reported in \cite{dromundo2022}, and vary (i) the size of the training sample used for cost estimation, (ii) the regularization parameter, and (iii) the strength of complementarities in the job finding rate function. Our results indicate that under reasonable levels of complementarities, and for realistic training sample sizes given the administrative data available, our method could improve (at virtually no cost) job finding rates by about 1 percentage point---an order of magnitude similar to very costly job search counseling interventions (e.g., \citet{behaghel2014}).

\subsection{Related literature}

The statistical treatment choice literature initiated by \cite{Manski2004} and \cite{Dehejia2005} has been a growing area of econometric research. Given a planner's welfare objective, there are multiple approaches to learning treatment choice rules from data. These include solving the statistical decision problem in finite samples \citep*{stoye2009minimax,stoye2012minimax, tetenov2012statistical, ishihara2021, yata2021, montielolea2023decision, kitagawa2022treatment}, solving for asymptotically optimal rules in limit experiments \citep*{HiranoPorter2009, christensen2025optimaldecisionrulespayoffs, masten2023minimax}, and performing empirical welfare maximization \citep*{kitagawa2018,KT19,Athey2021,MT17,sun2021empirical,Viviano21,KST21,Sakaguchi21,KLQ2025regret_aversion}. These works generally assume that a planner allocates a finite number of treatment arms over infinitely many individuals belonging to a superpopulation.\footnote{\cite*{Kallus_Zhou_2018}and \cite*{Ai_Fang_Xie2024} study policy learning for the allocation of a continuous treatment.} In statistical policy choice for two-sided matching problems, in contrast, the number of treatment arms (e.g., caseworkers) is as large as the number of individuals (e.g., job seekers) and each treatment arm has a tight capacity constraint (e.g., each caseworker can assist only a certain number of job seekers per month). In this setting, how can an optimal matching policy be learned from the data and its welfare regret assessed? This paper exploits recent advances in the empirical optimal transport literature to answer these questions.

The statistical properties of empirical optimal transport have been extensively studied in the case where the cost function is known. See \cite*{Genevay2019}, \cite{Rigollet2022}, and the references therein. On the other hand, the unknown cost function case is less studied. \cite*{Hundrieser2023} considers estimation and inference for the value of optimal transport with an estimated cost function, and obtains a distributional approximation for asymptotically valid inference. In contrast, we focus on a non-asymptotic regret bound for the estimated matching policy applied to a finite number of individuals drawn from a superpopulation. We obtain a regret bound under the minimal regularity condition that the cost function is bounded, and this bound is valid for any size of training sample and any number of individuals to be matched.

Optimal transport methods have been applied to many economic and econometric problems. See \cite{Galichon2016} for a recent monograph on the topic. These include the identification and estimation of Hedonic models \citep*{Ekeland_etal_2004}, the identification and estimation of a matching market surplus \citep*{Dupuy_Galichon_2014,Galichon_Salanie_2022}, partial identification \citep*{Galichon_Henry_2011,lei2025}, multivariate quantile analysis \citep*{Chernozhukov_etal_2017}, measurement errors and data combination \citep*{Schennach_Starck_2022, DHaultfoeuille_etal_2025}, and causal inference \citep*{Gunsilius_2023, pouliot2025}. The class of two-sided matching problems considered in this paper corresponds to the original Monge-Kantorovich optimal transport problem. \cite{bhattacharya2009} studies estimation of optimal matching policies with an estimated cost functions and discrete characteristics. \cite*{graham2011} and \cite*{graham2014} model the two-sided matching problem as choosing a copula between distributions of single indices aggregating multi-dimensional characteristics of each side, and perform estimation and inference for the optimal index coefficients as well as copula parameters. The optimal transportation approach has several advantages over the copula approach. First, the optimal transportation approach can accommodate multi-dimensional characteristics flexibly without reducing the characteristics of each side to a single index. Second, the optimal transportation approach is computationally attractive, as the estimation of an optimal matching policy can be reduced to convex optimization. 

The analytical and computational tools of optimal transport have been utilized in the treatment choice literature. \cite*{adjaho2022externally} and \cite*{Kido2025} model the difference between the sample and target populations by their Wasserstein distance, and study how treatment choice can be made robust to failures of external validity. Generalizing the analysis of \cite*{BhattacharyaDupas2012} with a binary treatment to the multi-valued treatment setting, \cite*{Sunada_Izumi_2025} formulates treatment assignment under capacity constraints as an optimal transportation problem and studies the local asymptotic optimality of an estimated optimal transport map in limit experiments. 


\section{Framework}

\subsection{The planner's assignment problem}

Consider a planner who matches one group of individuals $\mathcal{I} := \{1, \dots, n \}$ (e.g. job seekers) with another $\mathcal{J} := \{ 1, \dots, n\}$  (e.g. caseworkers). 
Matching is one-to-one and individuals are indivisible. Each individual $i \in \mathcal{I}$ is matched with only one individual $j \in \mathcal{J}$ and vice versa. 
Without loss of generality, we let the two sides consist of the same number of individuals, $|\mathcal{I}| = |\mathcal{J}| = n$. For example, if $|\mathcal{I}| > |\mathcal{J}|$, we can construct $\tilde{\mathcal{J}}$ such that $|\mathcal{I}| = |\tilde{\mathcal{J}}|$ by adding dummy individuals to $\mathcal{J}$. Any individual $i \in \mathcal{I}$ paired with a dummy individual in $\tilde{\mathcal{J}}$ is then an individual with no match in $\mathcal{J}$.

Before matching $\mathcal{I}$ and $\mathcal{J}$, the planner observes the characteristics of all individuals on both sides. Let
$\mathbf{X}^n \equiv (X_i: i \in \mathcal{I})$, $X_i \in \mathcal{X} \subset \mathbb{R}^{d_x}$, be the characteristics of the $n$ individuals in $\mathcal{I}$. For instance, if $\mathcal{I}$ is a pool of job seekers, $X_i$ could include $i$'s education level, previous earnings, work experience, an index of the risk of long-term unemployment, etc. We denote the empirical distribution on $\mathcal{X}$ constructed upon $\mathbf{X}^n$ by $\mu_n$. Similarly, let
$\mathbf{W}^n \equiv (W_j: j \in \mathcal{J})$, $W_j \in \mathcal{W} \subset \mathbb{R}^{d_w}$ be the observable characteristics of the $n$ individuals in $\mathcal{J}$. If $\mathcal{J}$ is a pool of caseworkers providing job search assistance, $W_j$ could include caseworker $j$'s demographic characteristics and an estimated index of value-added.  We denote the empirical distribution on $\mathcal{W}$ constructed upon $\mathbf{W}^n$ by $\nu_n$.

Define an \textit{assignment} to be a permutation: $\sigma: \mathcal{I} \to \mathcal{J}$, yielding $n$ pairs $(X_i, W_{\sigma(i)}), \; i=1,\dots,n$. An assignment $\sigma$ corresponds to a planner's edict dictating who in $\mathcal{I}$ matches with whom in $\mathcal{J}$. Let $\pi_{\sigma}$ be a probability distribution over the set of permutations $\{ \sigma \}$. 

\subsection{The cost function and optimal assignment}

Following the conventions of the optimal transport framework, the outcome when $i$ is matched with $j$ is the cost of the match, $Y_i(j) \in \mathbb{R_+}$. 
If the natural measure of the outcome of a match is output rather than cost, this can be transformed to a cost measured in terms of lost output relative to some maximum potential output.  In the example of matching job seekers and caseworkers, we define $Y_i(j)$ to be an indicator that takes value $Y_i(j)=1$ if job seeker $i$ assisted by caseworker $j$ remains unemployed within 6 months of enrolling in the program and value $Y_i(j)=0$ otherwise. 

We interpret the cost measure $Y_i(j), (i,j) \in \mathcal{I} \times \mathcal{J}$ as a set of potential outcomes in the sense that for any $i \in \mathcal{I}$, exogenously changing their match from $j$ to $j'$ causally shifts individual $i$'s match cost from $Y_i(j)$ to $Y_i(j')$. We assume that the potential outcomes $(Y_i(j): (i,j) \in \mathcal{I} \times \mathcal{J})$ are random variables defined on a superpopulation, and admit the following decomposition:
\begin{equation*}
Y_i(j) = c(X_i, W_{j}) + \epsilon_{i}(j),
\end{equation*}
where $c(x,w) := E[Y_i(j)|X_i =x, W_j = w]$ is the average cost if an individual $i$ with observable characteristics $X_i =x$ is exogenously matched with an individual $j$ with observable characteristics $W_j = w$. The average cost function is anonymous in the sense that $c(\cdot, \cdot)$ depends on $i$ and $j$ only through the values of their observable characteristics $(X_i,W_j)$. 
The residual term $\epsilon_i(j)$ captures the effect of any unobservable characteristics of $i$ and $j$ on their match cost. 
By construction, $E[\epsilon_i(j)|X_i, W_j] = 0$. We assume no-interference in the following sense: for every $(i,j) \in \mathcal{I} \times \mathcal{J}$
\begin{equation} \label{Assum:no interference}
Y_{i}(j) \perp (\mathbf{X}^n, \mathbf{W}^n) |X_i, W_j.
\end{equation}
That is, the match cost for $i$ and $j$ is statistically independent of the characteristics of all other individuals.\footnote{Another implicit no-interference assumption is embedded in the notation $Y_i(j)$ for denoting the (potential) outcome when $i$ is matched with $j$---as it states that this outcome does not depend on the assignment of other individuals.}

We define the planner's objective function to be the expected total cost of all $n$ matches given the profiles of the observable characterstics for both sides, $\mathbf{X}^n$ and $\mathbf{W}^n$. Under the no-interference assumption (\ref{Assum:no interference}), the expected total cost of assignment $\sigma$ is given by 

\begin{equation*}
E \left[ \sum_{i=1}^n Y_i(\sigma(i))  | \mathbf{X}^n, \mathbf{W}^n \right] =  \sum_{i=1}^n c(X_i,W_{\sigma (i)}).
\end{equation*}
If the planner generates an assignment $\sigma \sim \pi_{\sigma}$, the expected cost taking into account the randomness of assignment is given by
\begin{equation} \label{eq:expected cost with pi_sigma}
E_{\sigma \sim \pi_{\sigma}} \left[ \sum_{i=1}^n c(X_i,W_{\sigma(i)}) \right]
\end{equation}

We define the planner’s optimal assignment rule to be:
\begin{equation} \label{eq:pi_sigma*}
\pi_{\sigma}^{\ast} \in \arg \min_{\pi_{\sigma} \in \Pi_{\sigma}} E_{\sigma \sim \pi_{\sigma}} \left[ \sum_{i=1}^n c(X_i,W_{\sigma(i)}) \right],
\end{equation}
where $\Pi_{\sigma}$ is the set of distributions over the permutations of $n$ elements. 

Since the number of permutations is $n!$, the number of support points of $\pi_{\sigma}$ grows rapidly with $n$, and the optimization problem (\ref{eq:pi_sigma*}) is infeasible for all but very low values of $n$. 
To overcome this computational challenge, we employ Kantorovich's relaxation and transform (\ref{eq:pi_sigma*}) into the canonical Monge-Kantorovich optimal transport problem for the empirical distributions of $(\mathbf{X}^n,\mathbf{W}^n)$.

Let $\Pi(\mu_n,\nu_n)$ be the set of joint distributions on $\mathcal{X} \times \mathcal{W}$ whose marginal distribution on $\mathcal{X}$ coincides with $\mu_n$ and whose marginal distribution on $\mathcal{W}$ coincides with $\nu_n$. Notice that $\pi \in \Pi(\mu_n,\nu_n)$ corresponds to the probability masses $(\pi(X_i,W_j): (i,j) \in \mathcal{I} \times \mathcal{J})$ supported on $\mathbf{X}^n \times \mathbf{W}^n$.

Since the joint distribution on $\mathbf{X}^n \times \mathbf{W}^n$ representing permutation $\sigma$ belongs to $\Pi(\mu_n,\nu_n)$, any $\pi_{\sigma} \in \Pi_{\sigma}$ implies a unique $\pi \in \Pi(\mu_n, \nu_n)$. 
Conversely, by Birkhoff's theorem \citep{Birkhoff1946}, for any $\pi \in \Pi(\mu_n, \nu_n)$, there exists $\pi_\sigma \in \Pi_{\sigma}$ such that the joint distribution of $(X_i,W_{\sigma(i)}), \sigma \sim \pi_{\sigma}$ follows $\pi$. This implies that the search for an optimal assignment rule (\ref{eq:pi_sigma*}) can be reduced to the following discrete finite optimal transportation problem: 
\begin{equation} \label{eq:pi_n*}
\pi_n^{\ast} \in \arg \min_{\pi_n \in \Pi(\mu_n,\nu_n)} \pi_n(c), \mspace{15mu} \text{where} \mspace{10mu} \pi_n(c) = \sum_{i,j=1}^n c (X_i,W_j) \pi_n (X_i,W_j), \quad 
\end{equation}

\section{Estimation and implementation of optimal matching policies}

The planner's goal is to learn and implement the optimal matching rule $\pi_n^{\ast}$ defined in (\ref{eq:pi_n*}). However, in many applications, the average cost function $c(x,w)$ is unknown. This presents a fundamental challenge to learning the optimal policy. We consider estimating the cost function using training data and learning a matching policy based on the estimated cost function. To reduce the sensitivity of the estimated matching rule to estimation errors, we solve a sample analogue of the optimization problem (\ref{eq:pi_n*}) with a regularization term added to the objective function.   

\subsection{Estimation of the cost function}

The planner has access to training data consisting of $N \geq 1$ observed matches $(X_{\ell}, W_{\ell}: \ell=1, \dots, N)$ and the realized cost for each match $Y_{\ell}$, $\ell=1, \dots, N$. 
We assume that matches in the training data are randomized in the sense that conditional on the observable characteristics $X_{\ell}$ of the $\ell$-th individual, their match is assigned independently of their unobservable heterogeneity, denoted by $\epsilon_{\ell}$. 
This corresponds to the exogeneity condition considered in \cite{graham2011, graham2014},
\begin{equation}
W_{\ell} \perp \epsilon_{\ell} | X_{\ell}.
\end{equation}

Under this exogeneity assumption, the cost function $c(x,w)$ is identified as the conditional expectation function of $Y_{\ell}$ given the characteristics of the matched individuals $X_{\ell}$ and $W_{\ell}$. 

As a performance guarantee, we derive a welfare regret bound for an estimated matching policy which impose no assumptions on the cost function estimator. Let $\hat{c}(\cdot,\cdot)$ denote an estimator for $c(\cdot,\cdot)$. 
For the sake of later analysis, we define the $\mathcal{L}^1$ and $\mathcal{L}^2$ errors of $\hat{c}$ relative to the true cost function $c$ with respect to the product measure $\mu \otimes \nu$ (i.e., the independent coupling) of $\mu$ on $\mathcal{X}$ and $\nu$ on $ \mathcal{W}$:
\begin{align}
\| \hat{c} - c \|_{\mathcal{L}^1(\mu \otimes \nu)} &:= \int_{\mathcal{W}} \int_{\mathcal{X}} |\hat{c}(x,w) - c(x,w)| d\mu(x) d\nu(w), \\
\| \hat{c} - c \|_{\mathcal{L}^2(\mu \otimes \nu)} &:= \left( \int_{\mathcal{W}} \int_{\mathcal{X}} |\hat{c}(x,w) - c(x,w)|^2 d\mu(x) d\nu(w) \right)^{1/2}.
\end{align}

\subsection{Learning an optimal matching policy: entropy-regularized optimal transport}

Given the definition of the optimal matching policy (\ref{eq:pi_n*}), a natural approach is to replace the true cost function $c$ with its estimated counterpart $\hat{c}$, and maximize $\pi_n(\hat{c})$ with respect to $\pi_n \in \Pi (\mu_n, \nu_n)$. 

This plug-in approach faces several issues. 
First, since the optimization problem is a linear program, the solution will be an extreme point of the polyhedron constraint set $\Pi(\mu_n, \nu_n)$. 
As solutions of this type can be sensitive to perturbations in the objective function, an optimal matching policy estimated using a plug-in approach may be sensitive to estimation errors in $\hat{c}$. 
Second, although linear programming is a type of convex optimization, it scales poorly and this limits the scope of applications. 
For example, when matching $n=1000$ heterogeneous job seekers and case workers, the dimension of $\pi_n$ and the number of constraints are of the order $1000^2$, which is challenging even for modern linear programming solvers. 

Entropy regularized optimal transport (ROT) with iterative projection fitting (the Sinkhorn algorithm), as proposed in \cite{Cuturi2013} and \cite{Galichon_Salanie_2022}, mitigates these concerns.
Given a regularization parameter $1/\eta \geq 0$ and $(\mathbf{X}^n,\mathbf{W}^n)$, an empirical version of ROT considers the following optimization problem:
\begin{equation} \label{eq:empirical_primal_ROT}
\hat{\pi}_n^{ROT} \in \arg \min_{\pi_n \in \Pi(\mu_n,\nu_n)} \left\{  \pi_n(\hat{c})  + \frac{1}{\eta} KL(\pi_n \| \mu_n \otimes \nu_n) \right\},
\end{equation}
where $KL(\pi_n \| \mu_n \otimes \nu_n)$ is the Kullback-Leibler divergence of $\pi_n$ from the product empirical measure $\mu_n \otimes \nu_n$,
\begin{equation} 
KL(\pi \| \mu_n \otimes \nu_n) = \begin{cases} \sum_{i,j=1}^n \pi(X_i,W_j) \log\left( \frac{\pi(X_i,W_j)}{1/n^2} \right), & \text{if the support of $\pi$ is contained in $\mathbf{X}^n \times \mathbf{W}^n$.} \\
\infty, & \text{otherwise}
\end{cases} \notag
\end{equation}
Since $KL(\pi_n \| \mu_n \otimes \nu_n)$ increases as $\pi_n$ diverges from the independent coupling $\mu_n \otimes \nu_n$, the additional term penalizes $\pi_n$'s that concentrate at the extreme points of $\Pi(\mu_n, \nu_n)$. 
This pushes $\hat{\pi}_n^{ROT}$ away from deterministic policies corresponding to permutations in $\Pi_{\sigma}$ and towards stochastic matching policies. 
The regularization parameter $1/\eta$ governs how close $\hat{\pi}_n^{ROT}$ is to pure random matching. 
A smaller $\eta$ corresponds to a larger penalty, and forces $\hat{\pi}_n^{ROT}$ closer to random matching. As such, regularization makes $\hat{\pi}_n^{ROT}$ less sensitive to estimation errors in $\hat{c}$.
Conversely, as $\eta \to \infty$, the regularized solution $\hat{\pi}_n^{ROT}$ converges to the solution of the non-regularized empirical optimal transport problem. See, e.g., Proposition 4.1 in \cite{Peyre2020}.  

A computational advantage of ROT stems from the dual form of (\ref{eq:empirical_primal_ROT}). Let $f: \mathbf{X}^n \to \mathbb{R}$ and $g: \mathbf{W}^n \to \mathbb{R}$ be Lagrange multipliers for the equality constraints $\pi_n \in \Pi(\mu_n,\nu_n)$, and consider the Lagrangian of the primal problem:
\begin{align*}
& \min_{\pi_n}  \max_{f,g} \left\{ \pi_n(\hat{c})  + \frac{1}{\eta} KL(\pi_n \| \mu_n \otimes \nu_n) \right. \\ & \left. -\sum_{i=1}^n \left[ f(X_i) \left( \sum_{j=1}^n \pi_n(X_i,W_j) - \frac{1}{n} \right) \right] - \sum_{j=1}^n \left[ g(W_j) \left( \sum_{i=1}^n \pi_n(X_i,W_j) - \frac{1}{n} \right) \right] \right\}, 
\end{align*} 
Reversing the order of $\min_{\pi_n}$ and $\max_{f,g}$, we take the first-order conditions in $\pi_n$ given $(f,g)$ and obtain: 
\begin{equation*}
\pi_{n}(X_i,W_j) = \frac{1}{n^2} \cdot \exp\left( -\eta \hat{c}(X_i,W_j) + \eta f(X_i) + \eta g(W_j) \right).
\end{equation*}
Plugging this first-order condition into the Lagrangean yields the following dual form:
\begin{equation} \label{eq:empirical_dual_ROT}
(\hat{f}_n, \hat{g}_n) \in \arg \max_{f,g} \Big\{  \mu_n (f) + \nu_n(g) - \frac{1}{\eta} (\mu_n \otimes \nu_n) \left( e^{-\eta (\hat{c} - f - g) } -1 \right) \Big\},
\end{equation}
where $\hat{f}_n: \mathbf{X}^n \to \mathbb{R}$ and $\hat{g}_n: \mathbf{W}^n \to \mathbb{R}$ are optimal Lagrange multipliers in the dual problem, and $(\mu_n \otimes \nu_n)(\cdot)$ denotes integration with respect to $\mu_n \otimes \nu_n$.   

The dual problem is an unconstrained optimization problem with a concave objective function and two choice variables, $f$ and $g$, each of dimension $n$. The overall dimension of the choice variables in the dual problem is thus $2n$, whereas the dimension of the primal problem (\ref{eq:empirical_primal_ROT}) is $n^2$. Note that the dual objective function is invariant to $(f+a,g-a)$ for any translation vector $a \in \mathbb{R}^n$. Hence, with a location normalization of one of the vectors of choice variables such as 
\begin{equation}\label{eq:location normalization g}
\sum_{j=1}^n g(W_j) = 0, 
\end{equation}
the solution of the dual problem can be made unique. The main computational advantage of the formulation \eqref{eq:empirical_dual_ROT} is that it can be cast as a matrix scaling problem to which an iterative projection algorithm called the Sinkhorn algorithm can be applied. See, Proposition 4.3 in \cite{Peyre2020}.

With the solution $(\hat{f}_n,\hat{g}_n)$ in hand, we can express the optimal plan as, for each $i \in \mathcal{I}$ and $j \in \mathcal{J}$
\begin{equation} \label{eq:pi_hat_ROT}
\hat{\pi}_n^{ROT} (X_i,W_j) = \frac{1}{n^2} e^{-\eta (\hat{c}(X_i,W_j) - \hat{f}_n(X_i) - \hat{g}_n(W_j)) }.
\end{equation}
Note that $\hat{\pi}_n^{ROT}$ itself does not give pairings between $\mathcal{I}$ and $\mathcal{J}$. The last step is applying the Birkhoff–von Neumann algorithm to transform $\hat{\pi}_n^{ROT}$ into a distribution over permutations $\hat{\pi}_{\sigma}$. Matches are then formed by drawing $\sigma \sim \hat{\pi}_\sigma$. The average cost of the resulting pairs is $\hat{\pi}_n^{ROT}(c)$.  



\section{Performance guarantee}

Given $(\mu_n, \nu_n)$, we have built a recommended matching policy $\hat{\pi}_n^{ROT}$. This section assesses the statistical behavior of $\hat{\pi}_n^{ROT} (c)$ in terms of its average cost performance of with respect to the sampling distribution of the training data $(Y_{\ell}, X_{\ell}, W_{\ell}: \ell=1, \dots, N) \sim P^N$ and the characteristics of the individuals to be matched $(\mathbf{X}^n, \mathbf{W}^n)$. Specifically, we derive a non-asymptotic upper bound for the regret of the average cost of $\hat{\pi}_n^{ROT}$ and characterize its convergence rate.  

For this goal, define the oracle ROT matching policy given $(\mu_n,\nu_n)$ with knowledge of the true cost function $c(x,y)$ to be
\begin{equation}
\pi_n^{ROT} \in \arg \min_{\pi_n \in \Pi(\mu_n, \nu_n)} \left\{ \pi_n(c) + \frac{1}{\eta} KL(\pi_n \| \mu_n \otimes \nu_n ) \right\}. 
\end{equation}
Let $\pi_n^{\ast}(c)$ be the global minimum average cost. This corresponds to the average cost attained by $\pi^{\ast}_n \in \arg \min_{\pi_n \in \Pi(\mu_n, \nu_n)} \left\{ \pi_n(c) \right\}$, the oracle optimal matching policy without regularization. Let the average cost attained by a feasible regularized matching policy be $\hat{\pi}_n^{ROT}(c)$. 
We can then define the expected regret of implementing policy $\hat{\pi}_n^{ROT}$ given $(\mu_n,\nu_n)$ relative to the oracle unregularized policy $\pi_n^{\ast}$ as
\begin{equation}
Re(\hat{\pi}_n^{ROT}) := \hat{\pi}_n^{ROT}(c) - \pi_n^{\ast}(c).
\end{equation}
In what follows, we obtain a uniform upper bound for $Re(\hat{\pi}_n^{ROT})$ and discuss how its average (over samples) depends on the size of the training sample, the regularization parameter $\eta$, and the size of the matching market $n$. 

Since the regularization term is nonnegative, we have the following inequality for $Re(\hat{\pi}_n^{ROT})$:
\begin{align}
Re(\hat{\pi}_n^{ROT}) \leq  & \hat{\pi}_n^{ROT}(c) + \frac{1}{\eta} KL(\hat{\pi}_n^{ROT} \| \mu_n \otimes \nu_n ) - \pi_n^{ROT}(c) - \frac{1}{\eta} KL(\pi_n^{ROT} \| \mu_n \otimes \nu_n ) \notag \\
& + \pi_n^{ROT}(c) + \frac{1}{\eta} KL( \pi_n^{ROT} \| \mu_n \otimes \nu_n ) - \pi_n^{\ast}(c) \notag \\
= & Re^{ROT}(\hat{\pi}_n^{ROT}) + \underbrace{\pi_n^{ROT}(c) + \frac{1}{\eta} KL( \pi_n^{ROT} \| \mu_n \otimes \nu_n ) - \pi_n^{\ast}(c) }_{\text{Regularization bias}}, \label{eq:regret_upper_bound0}
\end{align}
where $ Re^{ROT}(\hat{\pi}_n^{ROT})$ is regret defined in terms of the regularized objective function:
\begin{equation}
Re^{ROT}(\hat{\pi}_n^{ROT}) := \hat{\pi}_n^{ROT}(c) - \pi_n^{ROT}(c) + \frac{1}{\eta} \left[ KL(\hat{\pi}_n^{ROT} \| \mu_n \otimes \nu_n) - KL( \pi_n^{ROT} \| \mu_n \otimes \nu_n) \right].
\end{equation}

There are two sources of randomness in regret and its upper bound (\ref{eq:regret_upper_bound0}).
\begin{enumerate}
\item Randomness in $\mu_n$ and $\nu_n$. To consider the long-run or repeated performance of $\hat{\pi}_n$, we take the expectation with respect to the sampling distribution of 
\begin{equation*}
(X_1,\dots, X_n, W_1, \dots, W_n) \sim \mu^n \otimes \nu^n,
\end{equation*}
where $\mu$ and $\nu$ are the population marginal distributions of $X_i$ and $W_j$, respectively, and the observable characteristics of the $n$ individuals on each side are drawn iid from $\mu$ and $\nu$.  
We denote the expectation with respect to this sampling distribution by $E_{P^n}[\cdot]$.
\item Randomness in the training sample: $\hat{c}(x,y)$ is estimated using a set of training data consisting of $N$ iid observations $(Y_{\ell},X_{\ell},W_{\ell}: \ell=1,\dots, N)$. We denote the expectation with respect to the training sample by $P^N$.
\end{enumerate}

To bound the welfare regret, consider the following steps:
\begin{enumerate}
\item Given $\hat{c}$ and $(\mu_n, \nu_n)$, obtain a non-stochastic inequality for $Re^{ROT}(\hat{\pi}_n^{ROT})$, where an upper bound depends on $\hat{c}$ and the size of the matching market $n$. 

\item To account for statistical uncertainty related to the training data, take the expectation of the upper bound obtained in step 1 with respect to the training sample, $E_P^N[\cdot]$.

\item Finally, take the expectation of the regret upper bound obtained in step 2 with respect to the distribution of the characteristics of the individuals to be matched $E_{P^n}[\cdot]$. 
\end{enumerate}

Note that, depending on the application, it may not be necessary to consider the randomness of the characteristics of the sample of individuals to be matched. In this case, expected regret must account only for the randomness of the training sample, and step 3 can be omitted.   

Based on the decomposition of (\ref{eq:regret_upper_bound0}), we bound the regret of the regularized cost $Re^{ROT}(\hat{\pi}_n^{ROT}) $ and the regularization bias separately.

Consider bounding $Re^{ROT}(\hat{\pi}_n^{ROT})$ as follows:
\begin{align}
Re^{ROT}(\hat{\pi}_n^{ROT}) = & \hat{\pi}_n^{ROT}(c) - \hat{\pi}_n^{ROT}(\hat{c}) + \hat{\pi}_n^{ROT}(\hat{c}) - \pi_n^{ROT}(c) \\ & + \frac{1}{\eta} \left[ KL(\hat{\pi}_n^{ROT} \| \mu_n \otimes \nu_n) - KL( \pi_n^{ROT} \| \mu_n \otimes \nu_n) \right] \\
= & \hat{\pi}_n^{ROT}(c) - \hat{\pi}_n^{ROT}(\hat{c}) +\Phi_n(\hat{c},\hat{f}_n, \hat{g}_n) - \Phi_n(c,f_n, g_n), \\
\leq & \hat{\pi}_n^{ROT}(c) - \hat{\pi}_n^{ROT}(\hat{c}) + \Phi_n(\hat{c}, \hat{f}_n, \hat{g}_n) - \Phi_n(c, \hat{f}_n, \hat{g}_n) \label{eq:regret_upper_bound1} 
\end{align}
where
\begin{equation}
\Phi_n(c,f,g) = \mu_n(f) + \nu_n(g) - \frac{1}{\eta} (\mu_n \otimes \nu_n) \left(e^{ -\eta c(x,y) + \eta f(x) + \eta g(y)} - 1 \right),
\end{equation}
is the dual objective function of the regularized OT problem,  $(\hat{f}_n,\hat{g}_n) = \arg \sup_{f,g} \Phi_n(\hat{c},f,g)$, and $(f_n,g_n) = \arg \sup_{f,g} \Phi_n(c,f,g)$. Note that the inequality (\ref{eq:regret_upper_bound1}) holds as $(f_n,g_n)$ maximizes $\Phi_n(c,f,g)$, which implies that $\Phi_n(c,f_n,g_n) \geq \Phi_n(c,\hat{f}_n,\hat{g}_n)$.

We obtain an upper bound for $Re^{ROT}(\hat{\pi}_n^{ROT})$ by bounding each of the two difference terms in (\ref{eq:regret_upper_bound1}). 

We impose the following regularity condition.

\begin{Assumption} \label{assump:bounded_c}
The cost function and its estimator are nonnegative and uniformly bounded,
\begin{equation}
0 \leq c(x,y), \hat{c}(x,y) \leq \bar{c} < \infty.
\end{equation}
\end{Assumption}
In our application to a job search assistance program, the cost measure is the probability of unemployment, which satisfies this assumption with $\bar{c} = 1$. 

We first present the following lemma, which is borrowed from \cite{Rigollet2022}.

\begin{Lemma} \label{lemma:bounds_for_pfg}
Let
\begin{equation}
\hat{p}_n(x,y) = \frac{d \hat{\pi}_n^{ROT}}{d(\mu_n \otimes \nu_n)}(x,y) = \exp(-\eta \hat{c}(x,y) + \eta \hat{f}_n(x) + \eta \hat{g}_n(y)).
\end{equation}
Under Assumption \ref{assump:bounded_c}, we have, for all $i,j=1,\dots,n$,
\begin{align}
& -\bar{c} \leq \hat{f}_n(x_i), \hat{g}_n(y_j) \leq \bar{c},\\
& e^{-3\eta \bar{c}} \leq \hat{p}_n(x_i,y_j) \leq e^{2 \eta \bar{c}}.
\end{align}
\end{Lemma}

For completeness, we provide a proof of this lemma in Appendix \ref{app:proofs}. Using this lemma, we obtain the following proposition, which gives finite sample upper bounds for $Re^{ROT}(\hat{\pi}_n^{ROT})$.

\begin{Proposition} \label{prop:regret_ROT}
Under Assumption \ref{assump:bounded_c}, 
\begin{align} \label{eq:prop1_bound1}
 \left| \hat{\pi}_n^{ROT}(c(x,y)) - \hat{\pi}_n^{ROT}(\hat{c}(x,y)) \right| & \leq e^{2\eta \bar{c}} (\mu_n \otimes \nu_n)(| \hat{c}-c |), \\ \label{eq:prop1_bound2}
 \left| \Phi_n(\hat{c}, \hat{f}_n, \hat{g}_n) - \Phi_n(c, \hat{f}_n, \hat{g}_n)  \right| & \leq e^{2\eta \bar{c}} (\mu_n \otimes \nu_n)((c-\hat{c})^2). 
\end{align}
Hence, we obtain
\begin{align}
&Re^{ROT}(\hat{\pi}_n^{ROT}) \leq e^{2\eta \bar{c}} \left[ \|\hat{c}-c\|_{\mathcal{L}^1(\mu_n \otimes \nu_n)}  + \|\hat{c}- c\|^2_{\mathcal{L}^2(\mu_n \otimes \nu_n)} \right], \label{eq:prop1_bound3} \\
&E_{P^n}[Re^{ROT}(\hat{\pi}_n^{ROT})] \leq e^{2\eta \bar{c}} \left[ \|\hat{c}-c\|_{\mathcal{L}^1(\mu \otimes \nu)}  + \|\hat{c}- c\|^2_{\mathcal{L}^2(\mu \otimes \nu)} \right], \label{eq:prop1_bound4} \\
&E_{P^N} \left[ E_{P^n}[Re^{ROT}(\hat{\pi}_n^{ROT})] \right] \leq e^{2\eta \bar{c}} \cdot E_{P^N} \left[ \|\hat{c}-c\|_{\mathcal{L}^1(\mu \otimes \nu)}  + \|\hat{c}- c\|^2_{\mathcal{L}^2(\mu \otimes \nu)} \right]. \label{eq:prop1_bound5}
\end{align}
\end{Proposition}

\bigskip

This proposition shows that the bound (\ref{eq:regret_upper_bound1}) can be expressed as the sum of $L^1$-error and the squared $L^2$-error of the cost function estimator multiplied by the constant $e^{2 \eta \bar{c}}$. This constant comes from a uniform upper bound for $\hat{p}_n(x,y)$ implied by Lemma \ref{lemma:bounds_for_pfg}. If $\hat{c}$ is consistent, the squared-$L^2$ error term vanishes faster than the $L^1$ error term, and the leading term of the convergence rate is governed by the $L^1$ error of the cost function estimator. For instance, if $c(x,y)$ is parametric and the parameters can be estimated at $N^{-1/2}$-rate, then $E_{P^N} \left[ \|\hat{c}-c\|_{L^1(\mu \otimes \nu)} \right]  = O(N^{-1/2})$ and $E_{P^N} \left[ \|\hat{c}-c\|^2_{L^2(\mu \otimes \nu)} \right]  = O(N^{-1})$. 

The next proposition provides a uniform upper bound for the regularization bias defined in (\ref{eq:regret_upper_bound0}). 

\begin{Proposition} \label{prop:regularization_bias}
The regularization bias defined in (\ref{eq:regret_upper_bound0}) satisfies
\begin{equation}
\pi_n^{ROT}(c) + \frac{1}{\eta} KL( \pi_n^{ROT} \| \mu_n \otimes \nu_n ) - \pi_n^{\ast}(c) \leq \frac{\log n}{\eta}.
\end{equation}
\end{Proposition}

Combining Propositions \ref{prop:regret_ROT} and \ref{prop:regularization_bias}, the decomposition in (\ref{eq:regret_upper_bound0}) yields the following bounds:

\begin{Theorem}
Under Assumption 1, we have
\begin{align}
&Re(\hat{\pi}_n^{ROT}) \leq e^{2\eta \bar{c}} \left( \|\hat{c}-c\|_{L^1(\mu_n \otimes \nu_n)}  + \|\hat{c}- c\|^2_{L^2(\mu_n \otimes \nu_n)} \right) + \frac{\log n}{ \eta}, \\
&E_{P^n}[Re(\hat{\pi}_n^{ROT})] \leq e^{2\eta \bar{c}} \left( \|\hat{c}-c\|_{L^1(\mu \otimes \nu)}  + \|\hat{c}- c\|^2_{L^2(\mu \otimes \nu)} \right) + \frac{\log n}{ \eta }, \label{eq:prop2_bound4} \\
&E_{P^N} \left[ E_{P^n}[Re(\hat{\pi}_n^{ROT})] \right] \leq e^{2\eta \bar{c}} \cdot E_{P^N} \left[ \|\hat{c}-c\|_{L^1(\mu \otimes \nu)}  + \|\hat{c}- c\|^2_{L^2(\mu \otimes \nu)} \right] + \frac{\log n}{ \eta }. \label{eq:prop2_bound5}
\end{align}
\end{Theorem}


The first term in these bounds captures the variance component of regret. This grows exponentially with $\eta$ (i.e., as the regularization parameter $1/\eta$ decreases), but decreases as the size of the training sample $N$ increases and the integrated $L^1$ and $L^2$-errors of $\hat{c}$ shrink.

The second term in these bounds captures the regularization bias, which decreases as $\eta$ increases. Hence, the choice of $\eta$ exhibits a bias-variance trade-off. 

Suppose that the convergence rate of the integrated $L^1$ error is $N^{-\alpha}$, e.g., the parametric rate of $\alpha = 1/2$, the nonparametric rate of $\alpha = -1/3$, etc. Then, if $\eta$ is chosen to be proportional to $\frac{\tilde{\alpha}}{2 \bar{c}} \log (N)$ with $\tilde{\alpha} < \alpha$, the upper bounds in (\ref{eq:prop2_bound4}) and (\ref{eq:prop2_bound5}) converge to zero at the rate $\log n / \log N$. 

\section{Numerical Examples}

\subsection{Numerical example featuring positive assortative matching}\label{sec:numerical example 1}

This first numerical example is designed to be a simple data generating process within which we can explore the performance of ROT and unregularized OT plans learned using an estimated cost function. We evaluate how the performance---in terms of regret relative to the unregularized oracle optimal policy---of the ROT plan derived from $\hat{c}$ depends on (i) the size of the training sample and (ii) the value of the regularization parameter $1/\eta$.

\paragraph{Data generating process.} The characteristics of job seekers $X_{\ell}$ and caseworkers $W_{\ell}$ both consist of a single variable drawn independently from the uniform distribution on $[0,1]$.
    \[
    X_{\ell}, W_{\ell} \sim \mathcal{U}[0,1].
    \]
In the calibrated exercise of the next subsection, the job seekers' characteristic will correspond to a covariate-based prediction of their risk of remaining unemployed 6 months after entering unemployment, and the caseworkers' characteristic will correspond to their ability to place individuals in employment within 6 months of their registration as unemployed job seekers. The true job-finding probability, as a function of the job seeker's characteristic $x$ and their assigned caseworker's characteristic $w$. is given by:
    \[
    p(x, w) = \frac{1}{3}x^2 + \frac{1}{3}w^2 + \frac{1}{3}xw.
    \]
Since $\frac{\partial^2 p}{\partial x \partial w} > 0$, the optimal plan associated with this ``production'' function is positive assortative matching (PAM). The cost function used in the computation of the OT plans is $c(x,w) = 1-p(x,w) \in \left[0,1\right]$. The vector of outcomes for all pairs---a vector whose $\ell$th entry is a binary indicator taking value 1 if job seeker $\ell$ found a job within 6 months---is drawn as:
    \[
    Y_{\ell} \sim \mathcal{B}(1-c(X_{\ell}, W_{\ell})),
    \]
where $\mathcal{B}(p)$ denotes the Bernoulli distribution with success probability $p$.

\medskip

The training samples used to estimate the cost function are samples of size $N \in \{500,$ $5\ 000, 50\ 000, 500\ 000\}$ drawn from the DGP described above. The ``main'' sample of job seekers and caseworkers to be matched one-to-one consists of 100 job seekers and 100 caseworkers (slots). This is a rough approximation of the number of matches that have to be made every month in a given unemployment agency.\footnote{See footnote 6 in the next subsection for a back-of-the-envelop computation of the monthly number of matches to be made in a given unemployment agency.} In this example we do not explore sampling uncertainty coming from the observations to be matched. The characteristics of this sample are equally spaced over the support of $X$ and $W$, i.e., $[0,1]$.

\paragraph{Cost Function Estimation}

\begin{figure}[H]
    \centering

    \hspace{0.05\textwidth}
    \begin{subfigure}[t]{0.49\textwidth}
        \centering
        \includegraphics[width=\textwidth]{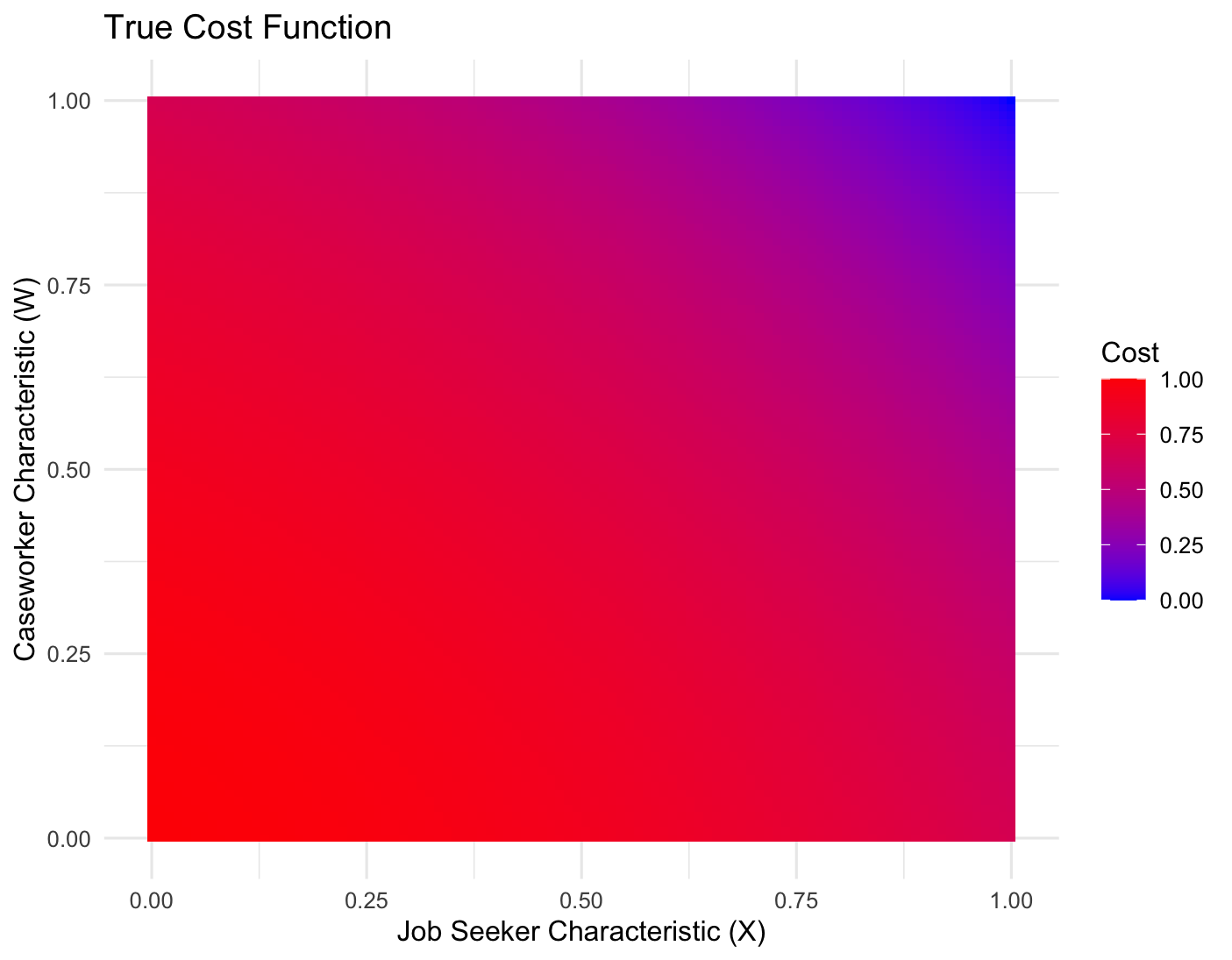}
        \caption{True cost $c(x, w)$}
        \label{fig:true_c_numex1}
    \end{subfigure}
    \hspace{0.05\textwidth}

    \vspace{0.5cm}

    \begin{subfigure}[t]{0.49\textwidth}
        \centering
        \includegraphics[width=\textwidth]{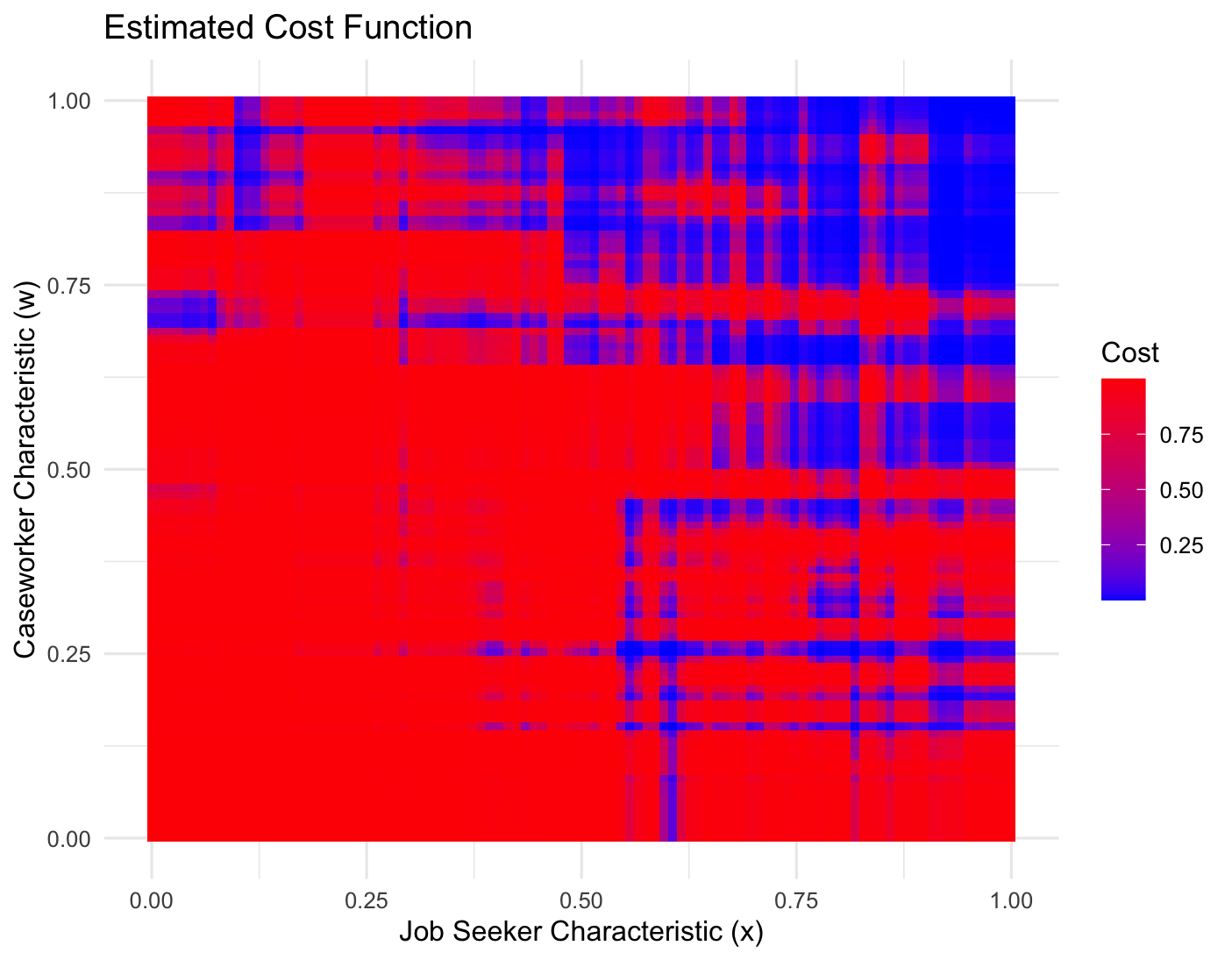}
        \caption{$\hat{c}(x,w)$, $N = 500$}
        \label{fig:hat_c_500}
    \end{subfigure}
    \begin{subfigure}[t]{0.49\textwidth}
        \centering
        \includegraphics[width=\textwidth]{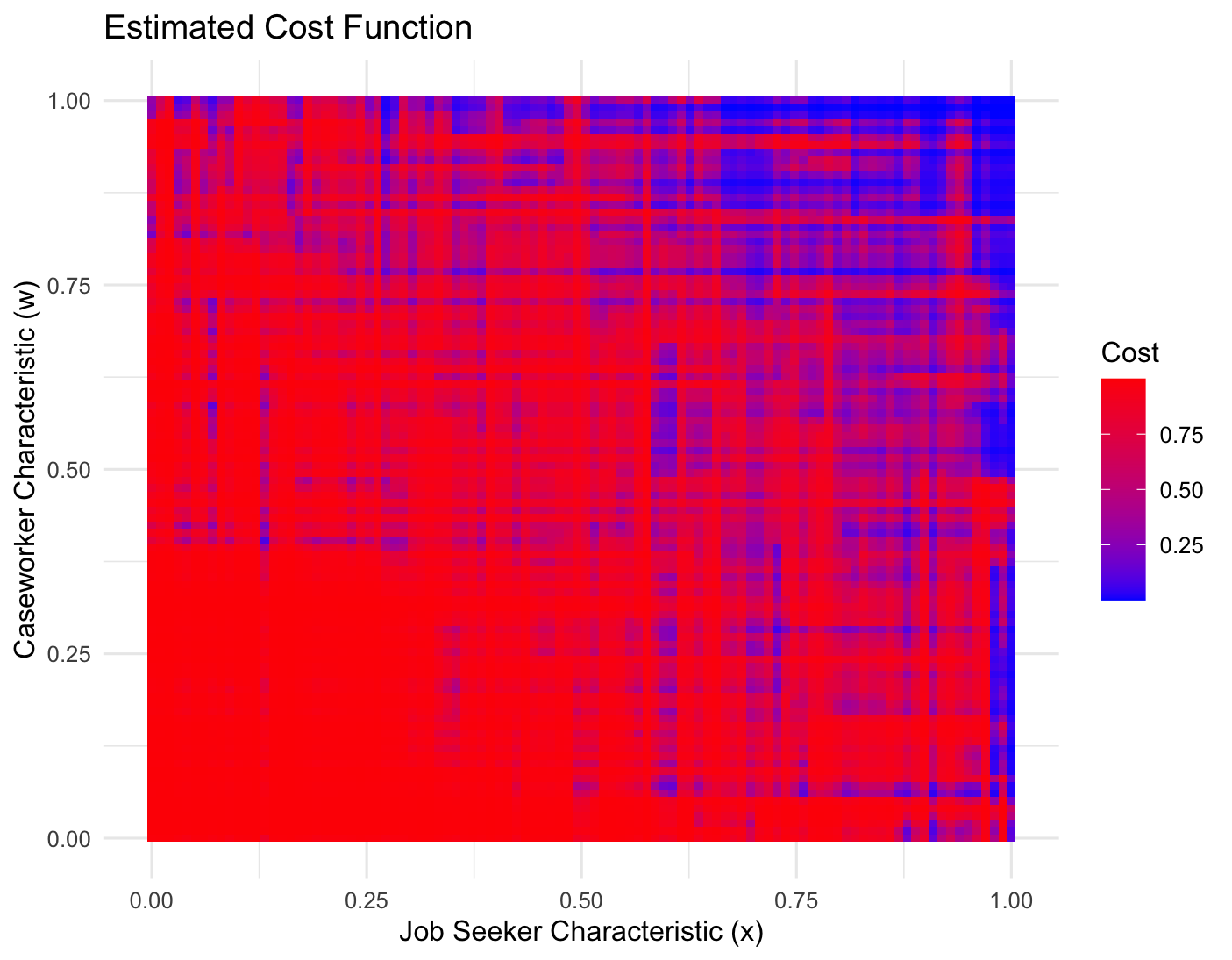}
        \caption{$\hat{c}(x,w)$, $N = 5,000$}
        \label{fig:hat_c_5000}
    \end{subfigure}

    \vspace{0.5cm}

    \begin{subfigure}[t]{0.49\textwidth}
        \centering
        \includegraphics[width=\textwidth]{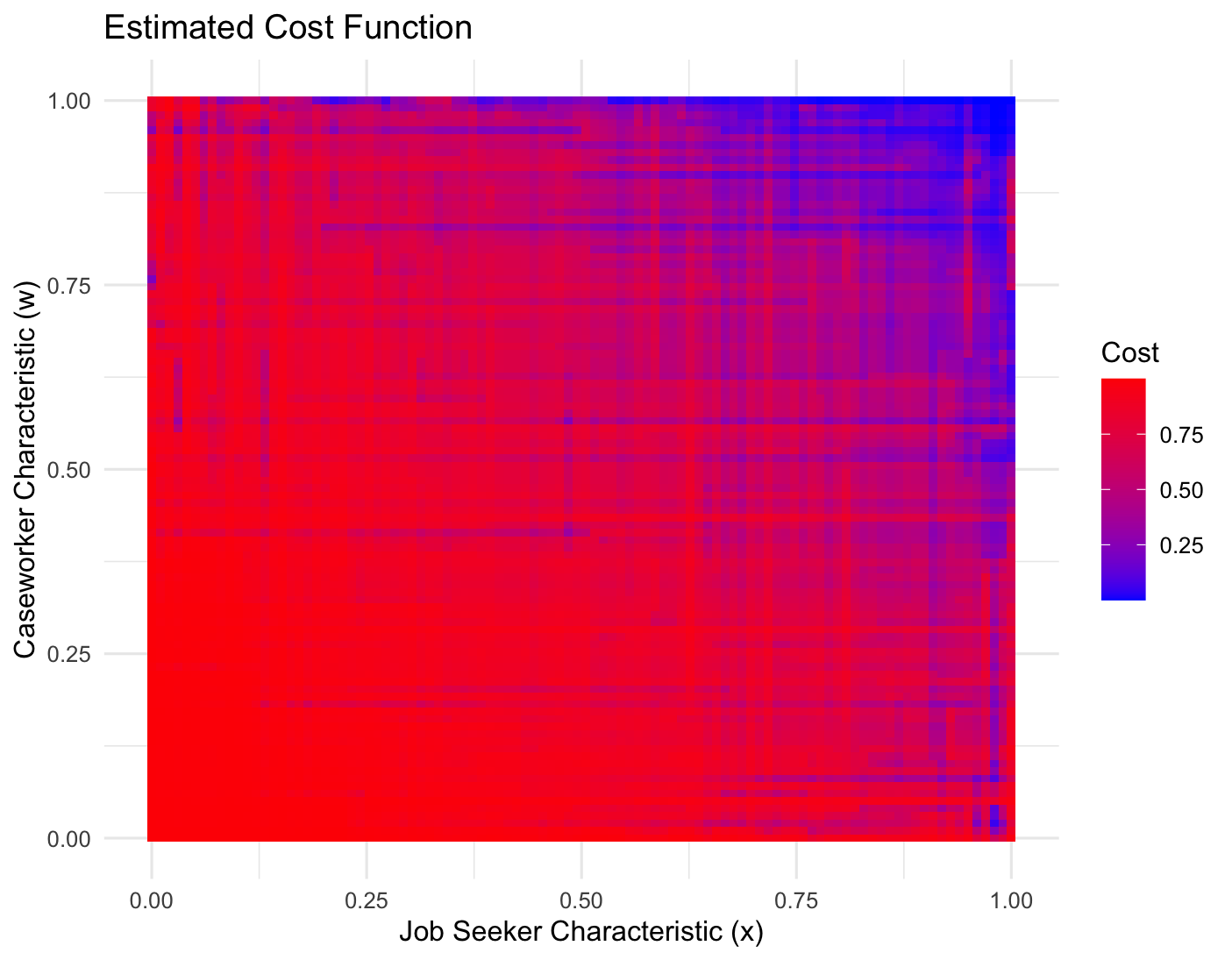}
        \caption{$\hat{c}(x,w)$, $N = 50,000$}
        \label{fig:hat_c_50000}
    \end{subfigure}
    \begin{subfigure}[t]{0.49\textwidth}
        \centering
        \includegraphics[width=\textwidth]{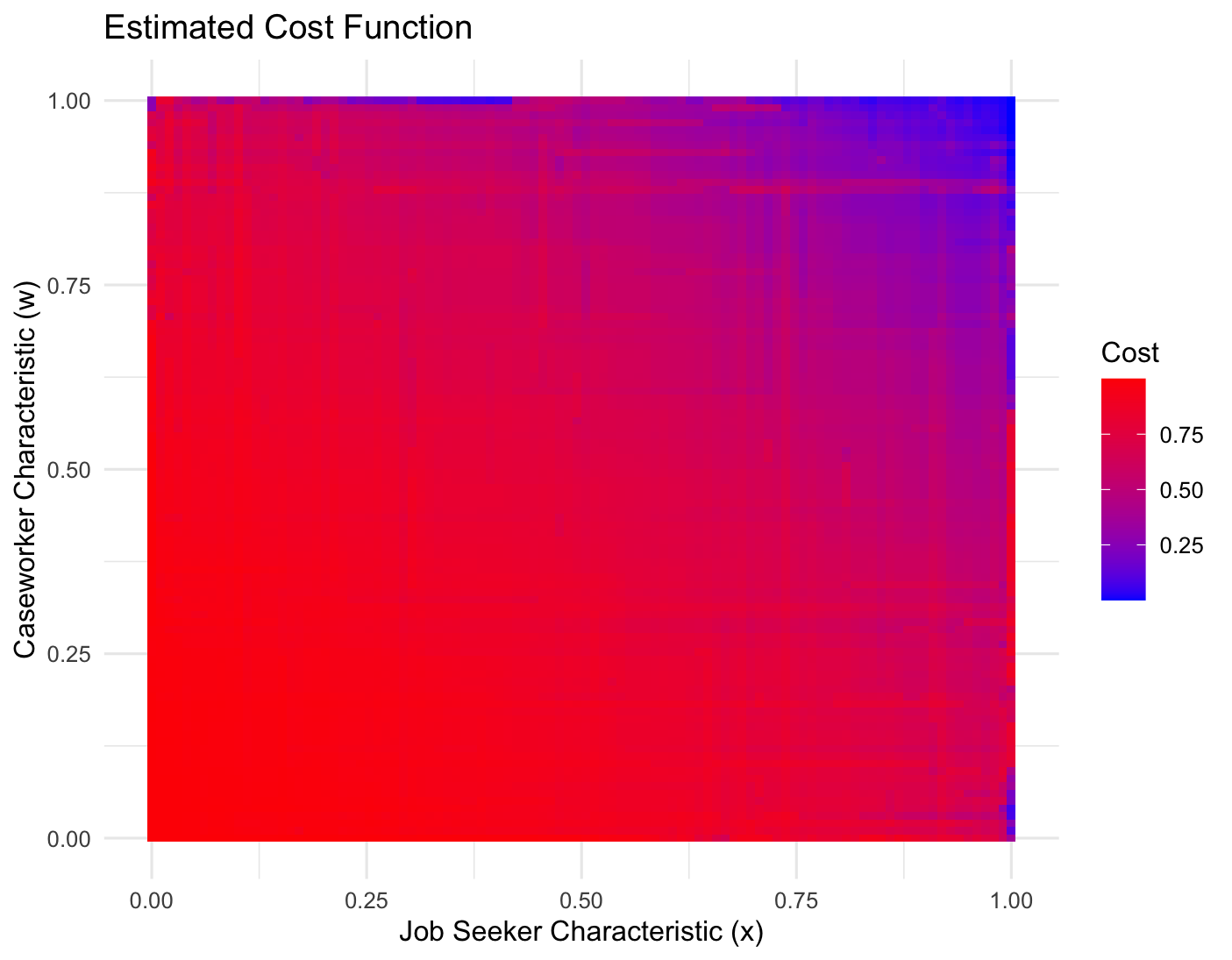}
        \caption{$\hat{c}(x,w)$, $N = 500,000$}
        \label{fig:hat_c_500000}
    \end{subfigure}

    \caption{Estimated cost functions $\hat{c}(x,w)$ for different training sample sizes, and the true cost function $c(x,w)$}
    \label{fig:cost_estimations}
\end{figure}

To obtain an estimated cost function $\hat{c}(x, w)$, we train an XGBoost model on the training sample, regressing $Y_\ell$ on $(X_\ell, W_\ell)$. Figure \ref{fig:cost_estimations} shows example estimated cost functions for different values of the training sample size $N$.

\paragraph{Welfare evaluation of the different transport plans}

We compare the average (over training samples) welfare generated by the following matching policies:

\begin{itemize}
    \item $\hat{\pi}^{\mathrm{ROT}}(c)$: ROT plan based on the estimated cost function $\hat{c}(x, w)$. This measure depends on the training sample size $N$.
    
    \item ${\pi}^{\mathrm{ROT}}(c)$: ROT plan computed using the true cost function $c(x, w)$.
    
    \item $\pi^*(c)$: Welfare from the optimal matching plan. By construction, for this DGP, this is positive assortative matching.
    
    \item $(\mu_n \otimes \nu_n)(c)$: Welfare under random matching.
\end{itemize}

The ROT plans are computed using the Sinkhorn algorithm in the log domain for numerical stability. We vary both the training sample size $N$ and the regularization parameter $1/\eta$.

\medskip

We report the welfare gain (or loss) of each ROT plan relative to random matching (the default policy in the current French Public Employment Services system), rescaled by the welfare gap between random matching and PAM:

\[
\frac{(\mu_n \otimes \nu_n)(c) - \pi_n^{\mathrm{ROT}}(c)}{(\mu_n \otimes \nu_n)(c) - \pi_n^*(c)} \quad \text{and} \quad \frac{(\mu_n \otimes \nu_n)(c) - \hat{\pi}_n^{\mathrm{ROT}}(c)}{(\mu_n \otimes \nu_n)(c) - \pi_n^*(c)}.
\]

This measure takes a negative value if the ROT plan performs worse than random matching, a value between 0 and 1 if the ROT plan outperforms random matching, and takes value 1 if the ROT plan coincides with unregularized OT plan. 

We choose to report rescaled welfare gains/losses as this numerical example is not calibrated in any way---as opposed to the one in the following subsection. Due to the computational burden associated with the estimation of the cost function, the quantities reported are averages over only 30 repetitions of the (i)-training sampling/(ii)-cost estimation/(iii)-transport plan computation procedure.

\medskip

\begin{figure}[H]
    \centering
    \includegraphics[width=0.8\textwidth]{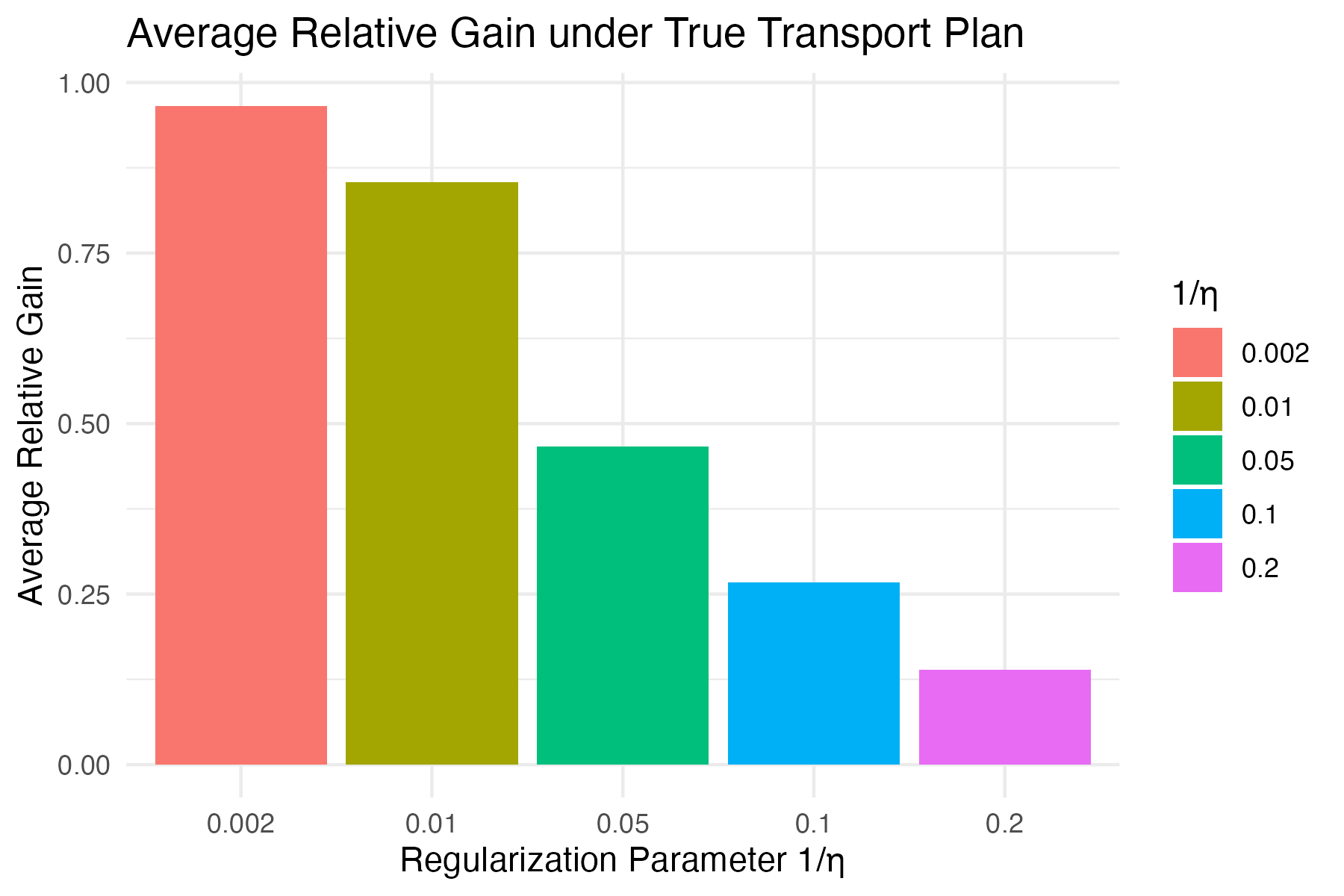}
    \caption{Relative welfare gain of the ROT plan learnt using the true cost $c(x,w)$ for different values of $1/\eta$} \label{fig:numex1_wgain_oracle}
\end{figure}

Figure \ref{fig:numex1_wgain_oracle} reports the average welfare gain of the oracle ROT plans learnt using the true cost function. As expected, as the regularization parameter goes to $0$, the average relative gain approaches that of the oracle unregularized plan.

\medskip

\begin{figure}[H]
    \centering
    \includegraphics[width=0.82\textwidth]{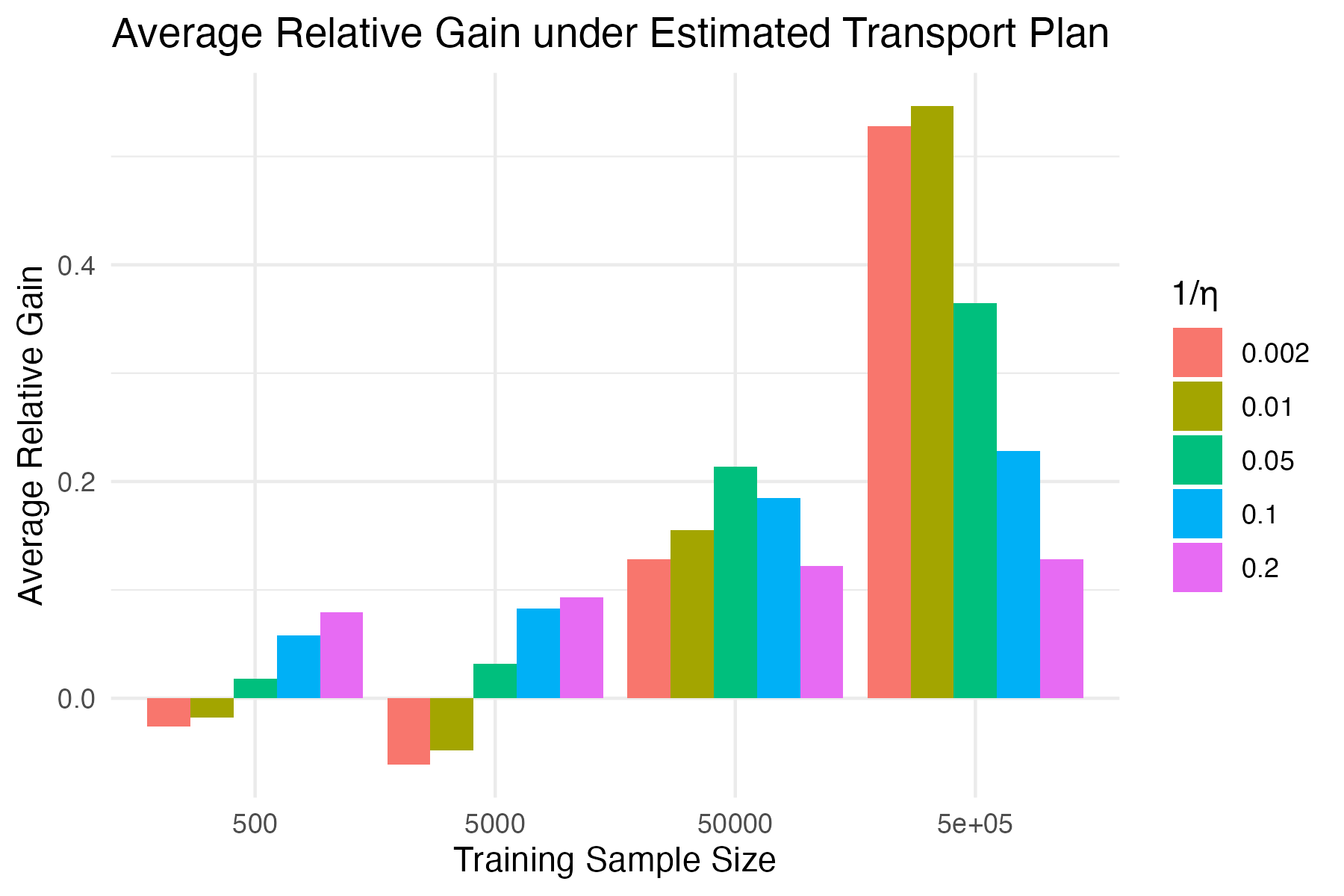}
    \caption{Relative welfare gain for ROT plan learnt using estimated cost $\hat{c}(x,w)$, depending on $N$ and $1/\eta$}\label{fig:numex1_wgain_ROTest}
\end{figure}

Figure \ref{fig:numex1_wgain_ROTest} reports the welfare gains associated with ROT plans learnt using an \textit{estimated} cost function. Interestingly, for the two smallest sample sizes, stronger regularization is beneficial in terms of welfare.

\subsection{Numerical example calibrated on French data}

To better understand the performance of our procedure in the context of the French Public Employment Services data,\footnote{We are currently constructing the dataset necessary for our analysis, under a data access agreement with the French Public Employment Services.} we calibrate our simulations using results on caseworker value added reported in \cite{dromundo2022}.

\begin{figure}[H]
    \centering
    \includegraphics[width=0.8\linewidth]{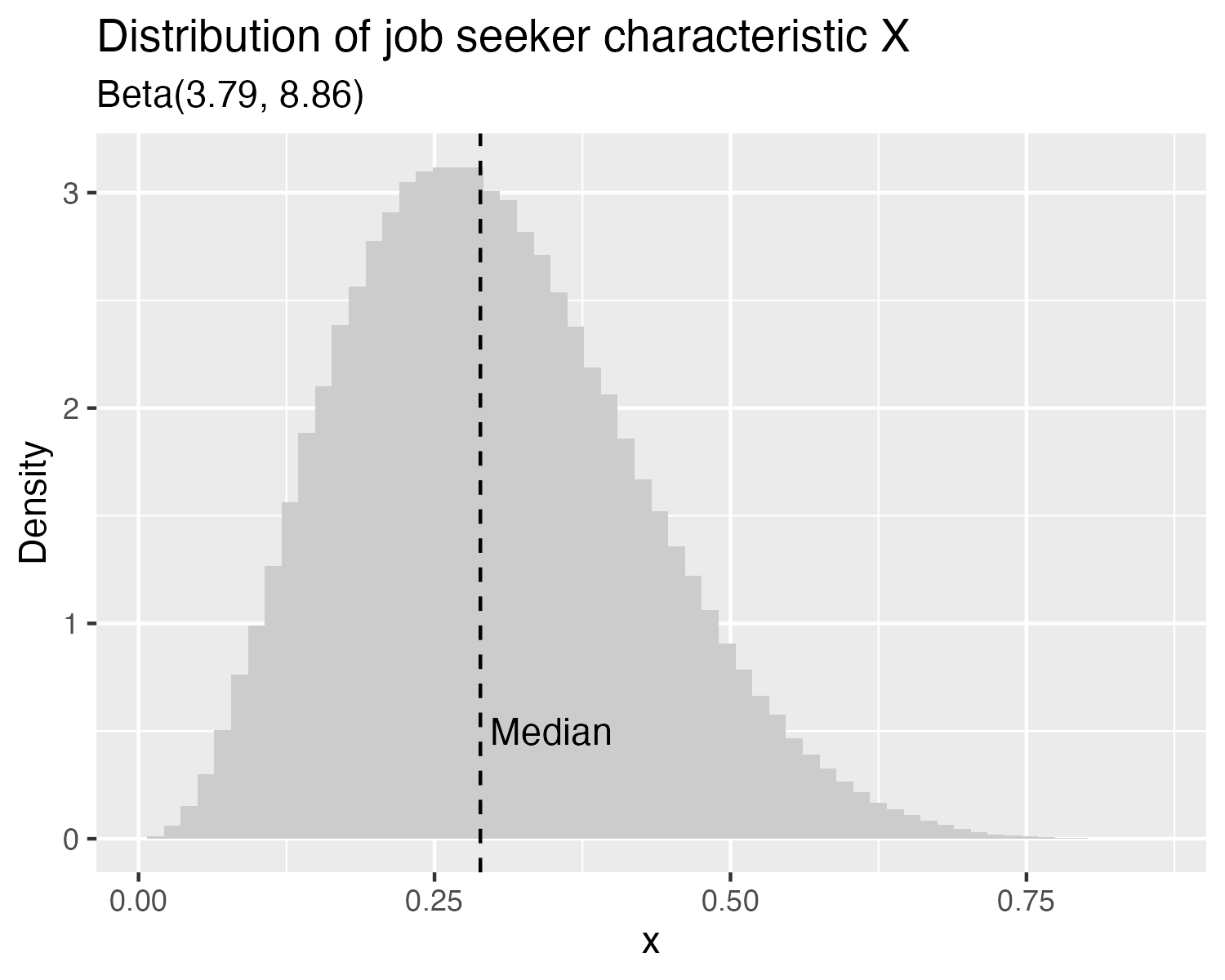}
    \caption{Distribution of the simulated job seekers's characteristic $X$}
    \label{fig:X_Beta_distrib}
\end{figure}

\paragraph{Data generating process.} A job seeker's characteristics consist of a single variable $X_{\ell}$ drawn from the $\operatorname{Beta}(\alpha = 3.7939, \beta=8.8634)$ distribution. $\alpha$ and $\beta$ are calibrated to match the moments of the distribution of the predicted 6-month job finding probability reported in Table 3 of \cite{dromundo2022}.\footnote{Specifically, we match the above and below median averages (respectively 0.20 and 0.40) of this predicted probability in the sample of job seekers studied in \cite{dromundo2022}.} Figure \ref{fig:X_Beta_distrib} shows the resulting distribution. A caseworker's characteristics consists of a single variable  $W_{\ell}$ drawn from the $\mathcal{N}(0,1)$ distribution. In Figure 1 of \cite{dromundo2022}, estimates of caseworkers' value added appear to follow a normal distribution.

\medskip

The true job-finding probability as a function of the job seeker's characteristic $x$ and their assigned caseworker's caracteristic $w$ is given by a standard logistic function:
    \[
    p(x, w) = \frac{\exp(a + b \cdot x + c \cdot  w + d \cdot x \cdot w)}{1 + \exp(a + b \cdot x + c \cdot  w + d \cdot x \cdot w)}
    \]
where the parameters $a$, $b$, $c$ and $d$ are calibrated to yield a given degree of complementarity. In particular, we define a gap $\gamma \in \{0.02, 0.06, 0.10\}$ in the effect of a one standard deviation increase in $w$ (from 0 to 1) on the job finding rate of job seekers with $x=0.20$ compared to job seekers with $x = 0.40$. We use the values $x = 0.20$ and $x = 0.40$ here as they are, respectively, the average values of $x$ for job seekers with below and above median characteristics. More precisely, we calibrate $(a,b,c,d)$ so that the function $p(x,w)$ matches
\begin{itemize}
    \item $p(x=0.20, w=0) = 0.20$: job seekers with a predicted 6-month job finding rate of $0.20$ have an average job finding rate of $0.20$ when matched with caseworkers with characteristic $w=0$.
    \item $p(x=0.40, w=0) = 0.40$: similar reasoning to above.
    \item $p(x=0.20, w=1) - p(x=0.20, w=0) = 0.02$: job seekers with a below median characteristic ($x=0.20$) benefit from a 2pp increase in 6-month job finding rate when matched with a one standard deviation above average caseworker ($w=1$ vs. an average caseworker $w=0$). This matches the ``marginal'' effect of $w$ for job seekers with a below median predicted job finding probability reported in \cite{dromundo2022}.
    \item $p(x=0.40, w=1) - p(x=0.40, w=0) - \left[p(x=0.20, w=1) - p(x=0.20, w=0)\right] \\= \gamma \in \{0.02, 0.06, 0.10\}$:  \cite{dromundo2022} find a statistically significant gap in the ``marginal'' effect of $w$ between above and below median job seekers of the order of magnitude of 1pp. Since the quantity they report is an average marginal effect among above and below median job seekers, it is likely that the target quantity here is larger than 1pp. In order to gauge the extent of complementarities that must exist in the data for our procedure to produce a meaningful increase in welfare, we allow it to vary between 2pp. and 10pp.
\end{itemize}
The cost function is given by $1-p(x,w) \in \left[0,1\right]$.

\medskip

The vector of outcomes ---a vector whose $\ell$th entry is a binary indicator taking value 1 if job seeker $\ell$ found a job within 6 months---is drawn as before as
    \[
    Y_{\ell} \sim \mathcal{B}(p(X_{\ell}, W_{\ell})),
    \]
where $\mathcal{B}(p)$ denotes the Bernoulli distribution with success probability $p$.

\medskip

Training samples---used to estimate the cost function---are of size $N \in \{500,$ $5\ 000, 50\ 000, 500\ 000, 5\ 000\ 000\}$ and follow the DGP described above. The 5 million observation sample is added to the sample sizes considered in Section \ref{sec:numerical example 1} as we expect the size of our training sample to be of this order of magnitude (several million observations).\footnote{Using the ILO definition, the stock of job seekers in France at any given point in time is roughly 5 million. We may have to filter our data before estimating the cost function, but we will have access to several years of data.} The ``main'' sample of job seekers and caseworkers to be matched one-to-one consists of 200 job seekers and 200 caseworkers (slots). This is a rough approximation of the number of matches that have to be made every month in a given unemployment agency.\footnote{\cite{dromundo2022} document that unemployment agencies in the Paris area have an average of 30 caseworkers, and on average a caseworker is assigned 30 job seekers each quarter. A ``caseworker'' in our DGP can be interpreted as one caseworker \textit{slot}, hence the number of job seekers determines the number of matches to be formed. In a typical Parisian agency, there are $30 \times 30 = 900$ matches a quarter, i.e., about $300$ per month. Parisian agencies are likely to be larger than the typical French agency. Therefore, in our numerical simulations, we match 100 or 200 job seekers and caseworkers (slots).} We do not explore the effect of sampling uncertainty coming from the observations to be matched. Instead, the job seekers to be matched are given characteristics corresponding to 200 quantiles of the calibrated Beta distribution, while caseworkers are given characteristics corresponding to 200 quantiles of a $N(0,1)$.

\paragraph{Cost Function Estimation}

\begin{figure}[H]
    \centering

    \begin{subfigure}[t]{0.49\textwidth}
        \centering
        \includegraphics[width=\textwidth]{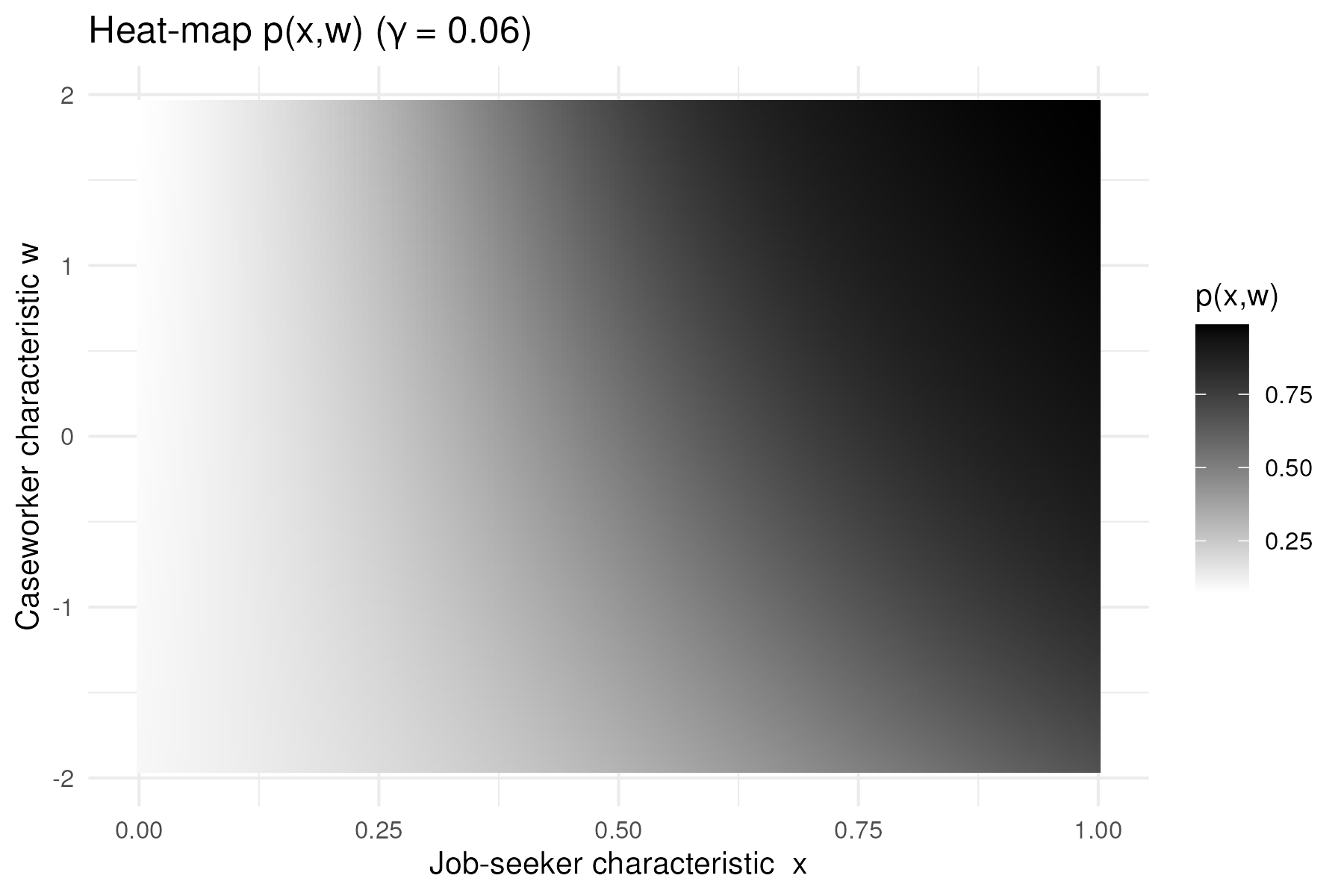}
        \caption{True cost $c(x, w)$}
        \label{fig:true_c}
    \end{subfigure}
    \begin{subfigure}[t]{0.49\textwidth}
        \centering
        \includegraphics[width=\textwidth]{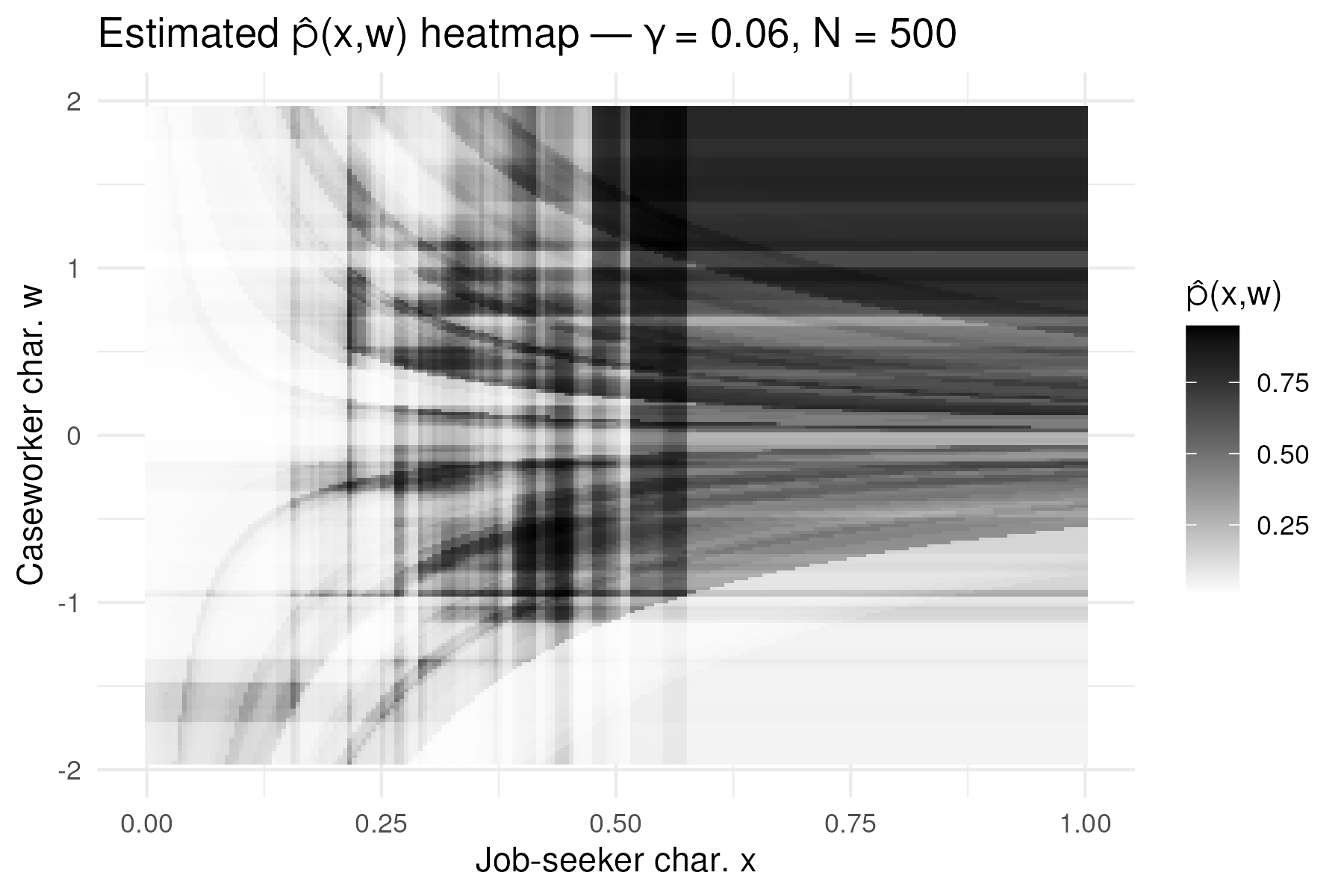}
        \caption{$\hat{c}(x,w)$, $N = 500$}
        \label{fig:numex2_hat_c_500}
    \end{subfigure}

    \vspace{0.5cm}

    \begin{subfigure}[t]{0.49\textwidth}
        \centering
        \includegraphics[width=\textwidth]{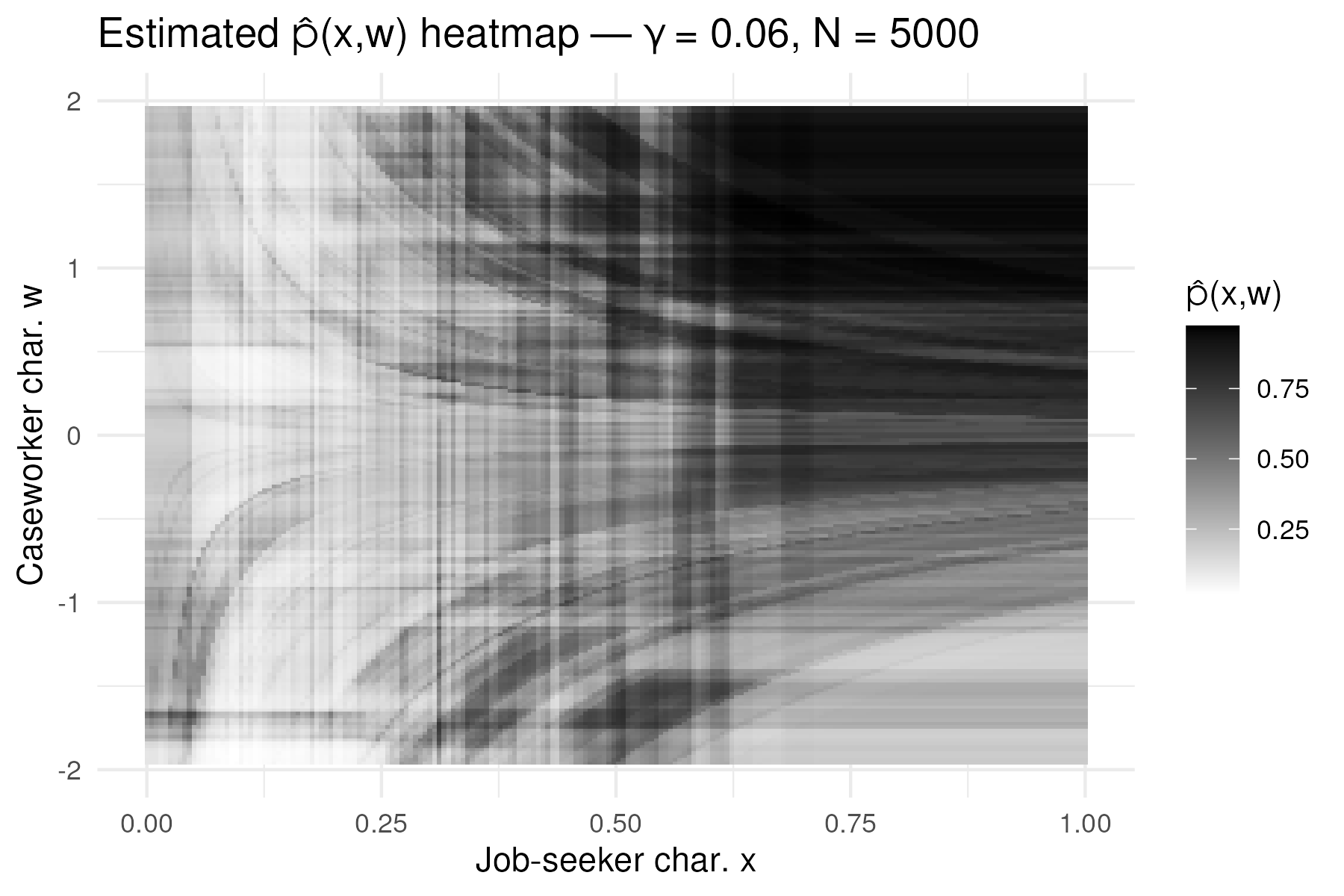}
        \caption{$\hat{c}(x,w)$, $N = 5,000$}
        \label{fig:numex2_hat_c_5000}
    \end{subfigure}
    \begin{subfigure}[t]{0.49\textwidth}
        \centering
        \includegraphics[width=\textwidth]{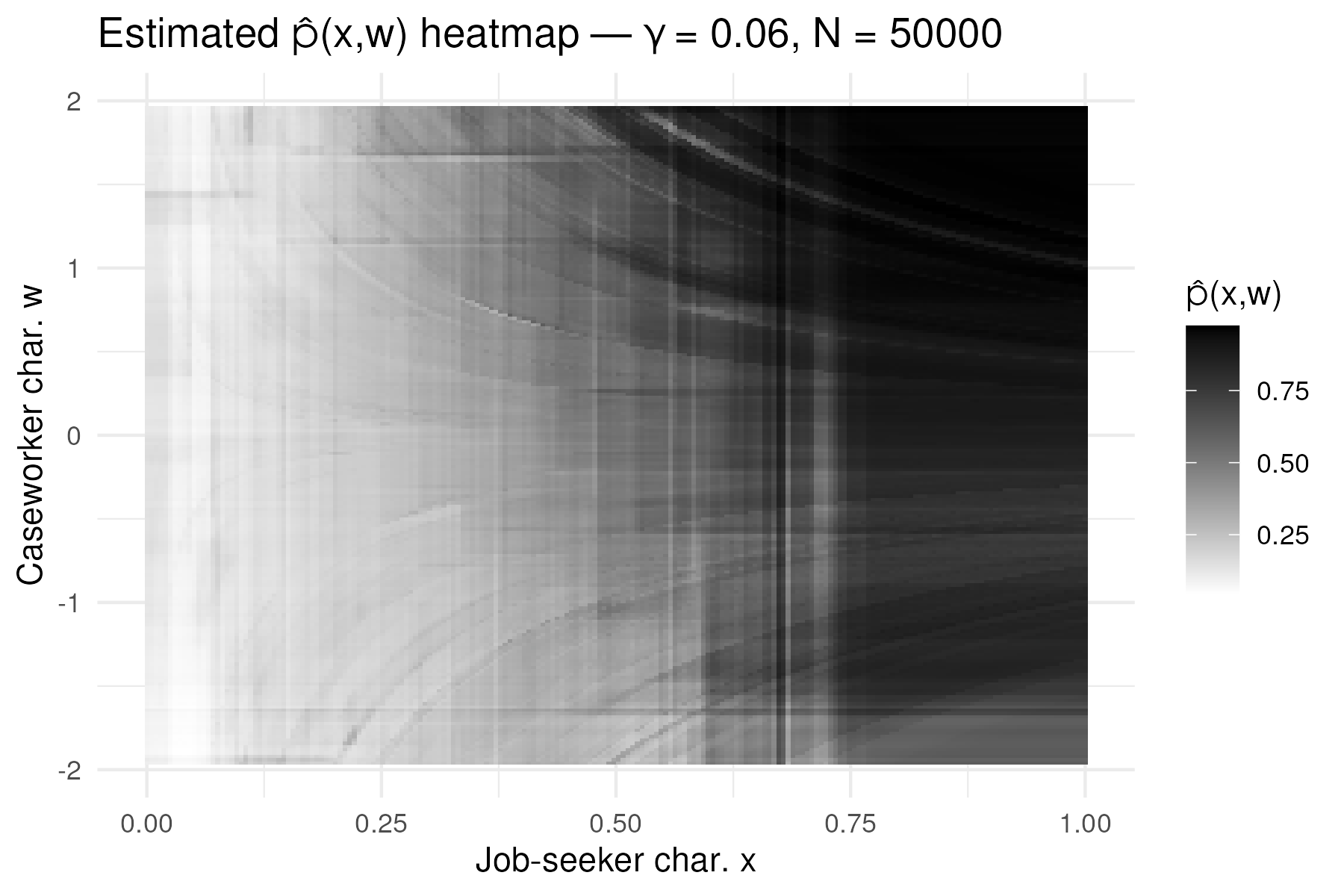}
        \caption{$\hat{c}(x,w)$, $N = 50,000$}
        \label{fig:numex2_hat_c_50000}
    \end{subfigure}
    \vspace{0.5cm}

    
    \begin{subfigure}[t]{0.49\textwidth}
        \centering
        \includegraphics[width=\textwidth]{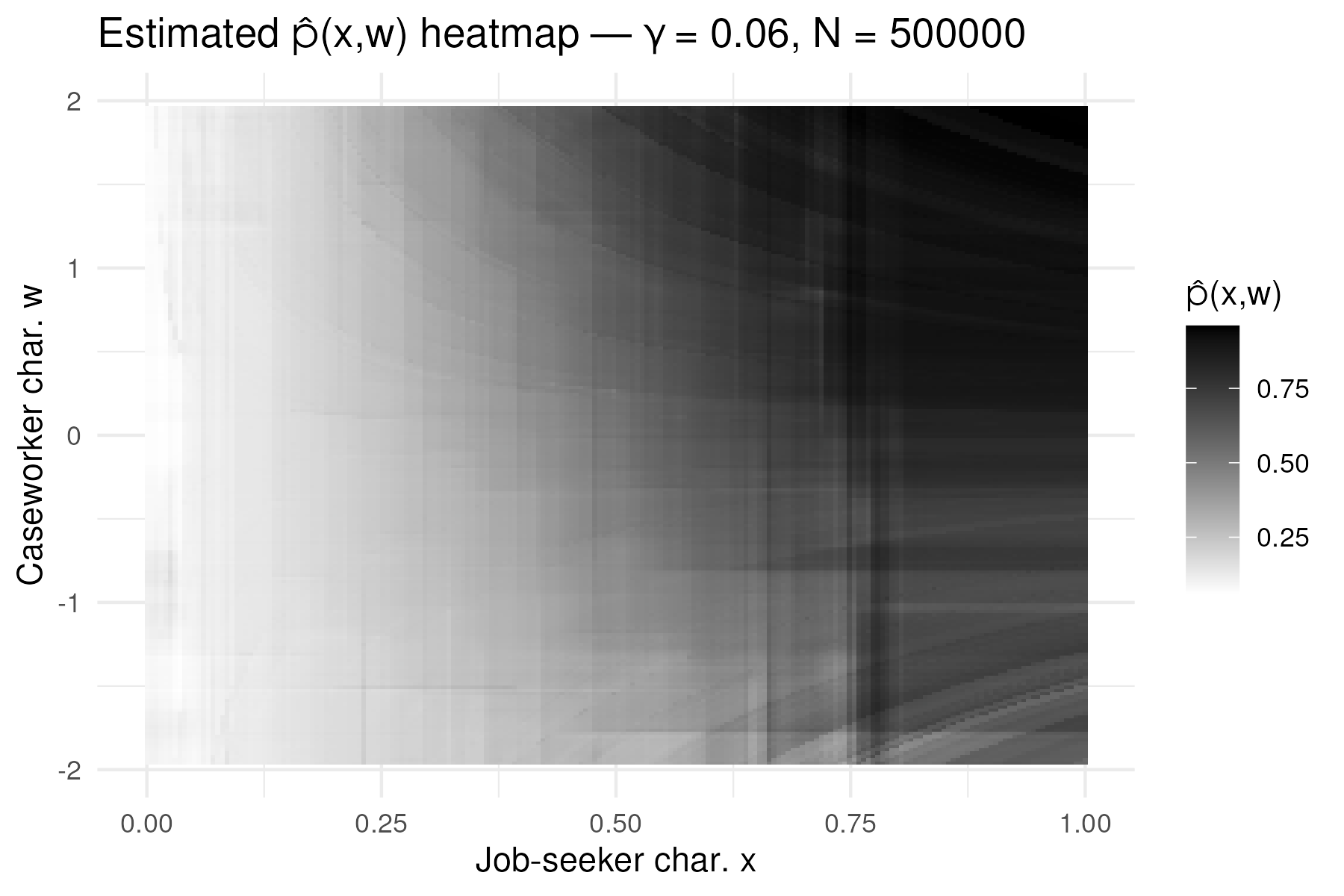}
        \caption{$\hat{c}(x,w)$, $N = 500,000$}
        \label{fig:numex2_hat_c_500000}
    \end{subfigure}
    \begin{subfigure}[t]{0.49\textwidth}
        \centering
        \includegraphics[width=\textwidth]{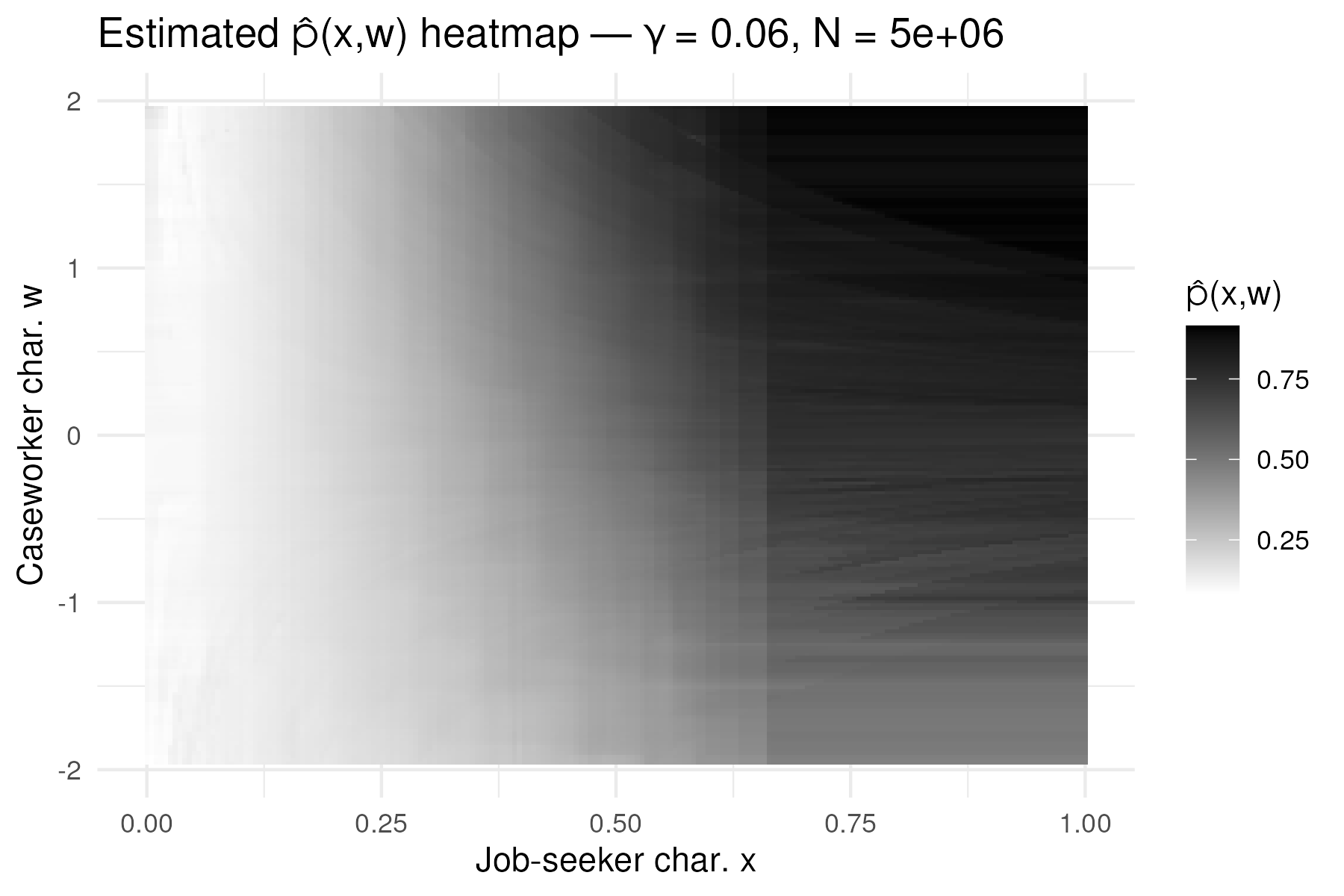}
        \caption{$\hat{c}(x,w)$, $N = 5,000,000$}
        \label{fig:numex2_hat_c_5000000}
    \end{subfigure}

    \caption{Estimated cost functions $\hat{c}(x,w)$ for different training sample sizes, and the true cost function $c(x,w)$ ($\gamma = 0.06$)}
    \label{fig:numex2_cost_estimations}
\end{figure}

\begin{figure}[H]
    \centering

    \begin{subfigure}[t]{0.49\textwidth}
        \centering
        \includegraphics[width=\textwidth]{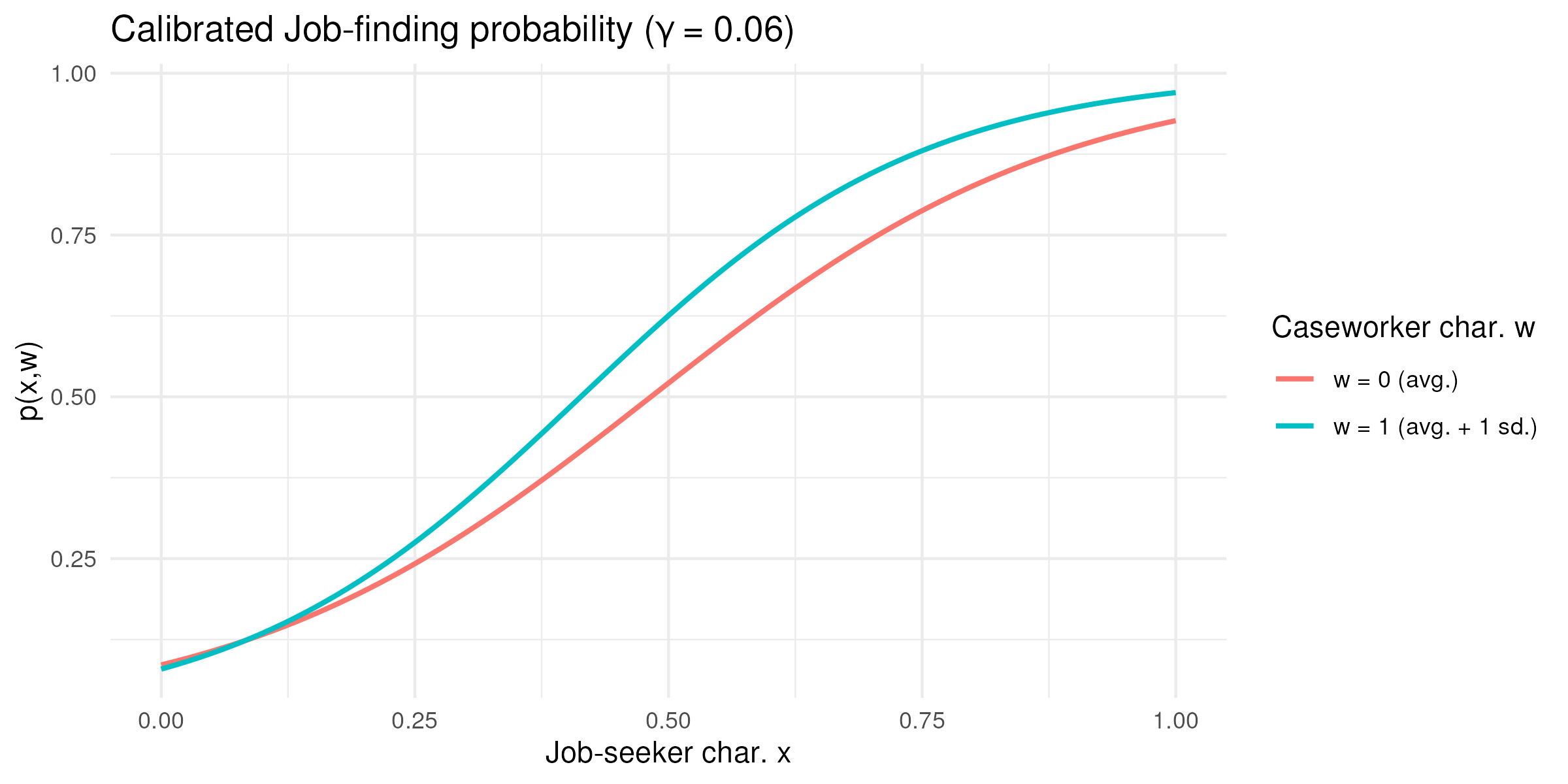}
        \caption{True cost $c(x, w)$}
        \label{fig:true_c_curve}
    \end{subfigure}
    \begin{subfigure}[t]{0.49\textwidth}
        \centering
        \includegraphics[width=\textwidth]{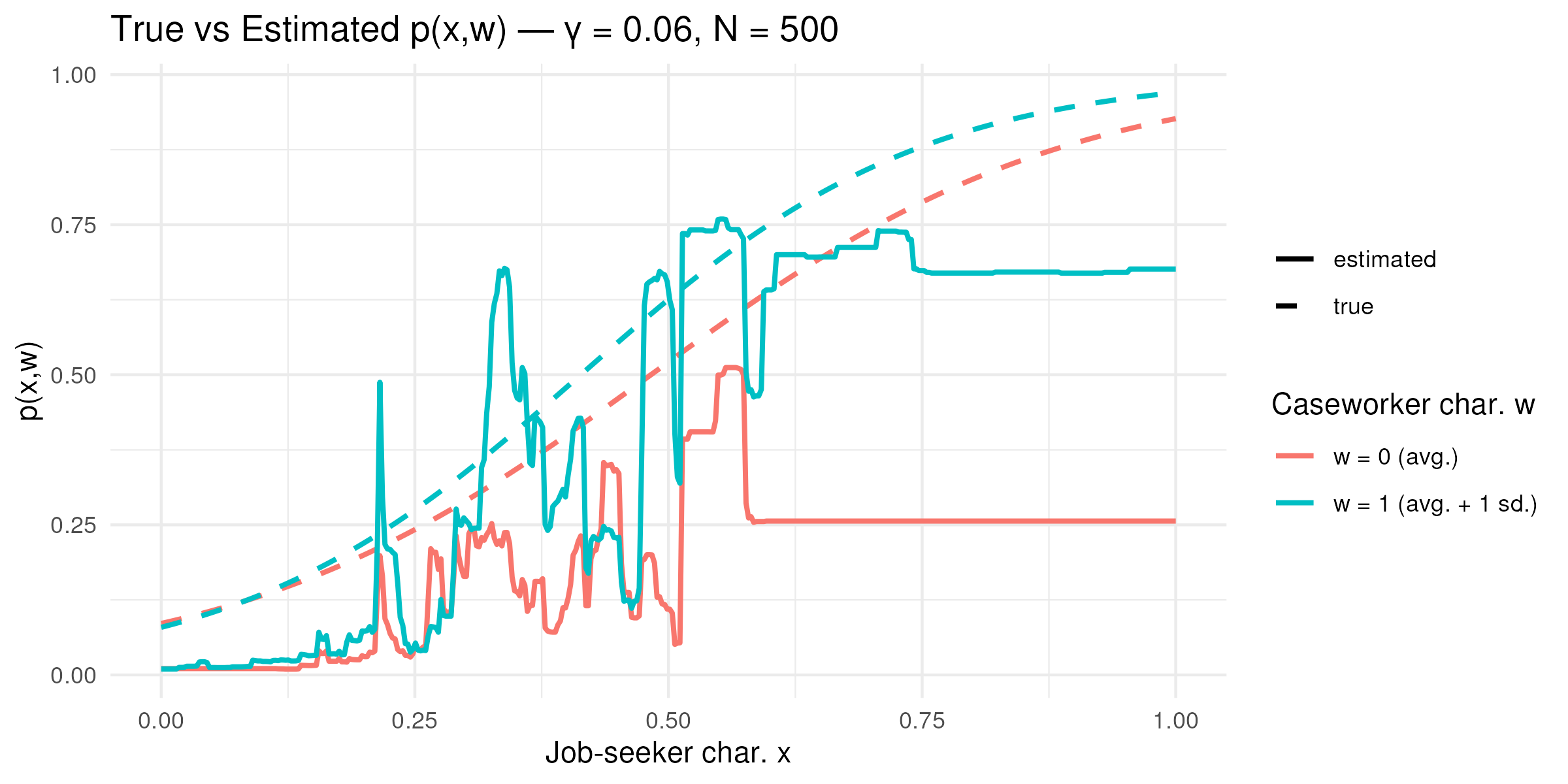}
        \caption{$\hat{c}(x,w)$, $N = 500$}
        \label{fig:numex2_hat_c_curve_500}
    \end{subfigure}

    \vspace{0.5cm}

    \begin{subfigure}[t]{0.49\textwidth}
        \centering
        \includegraphics[width=\textwidth]{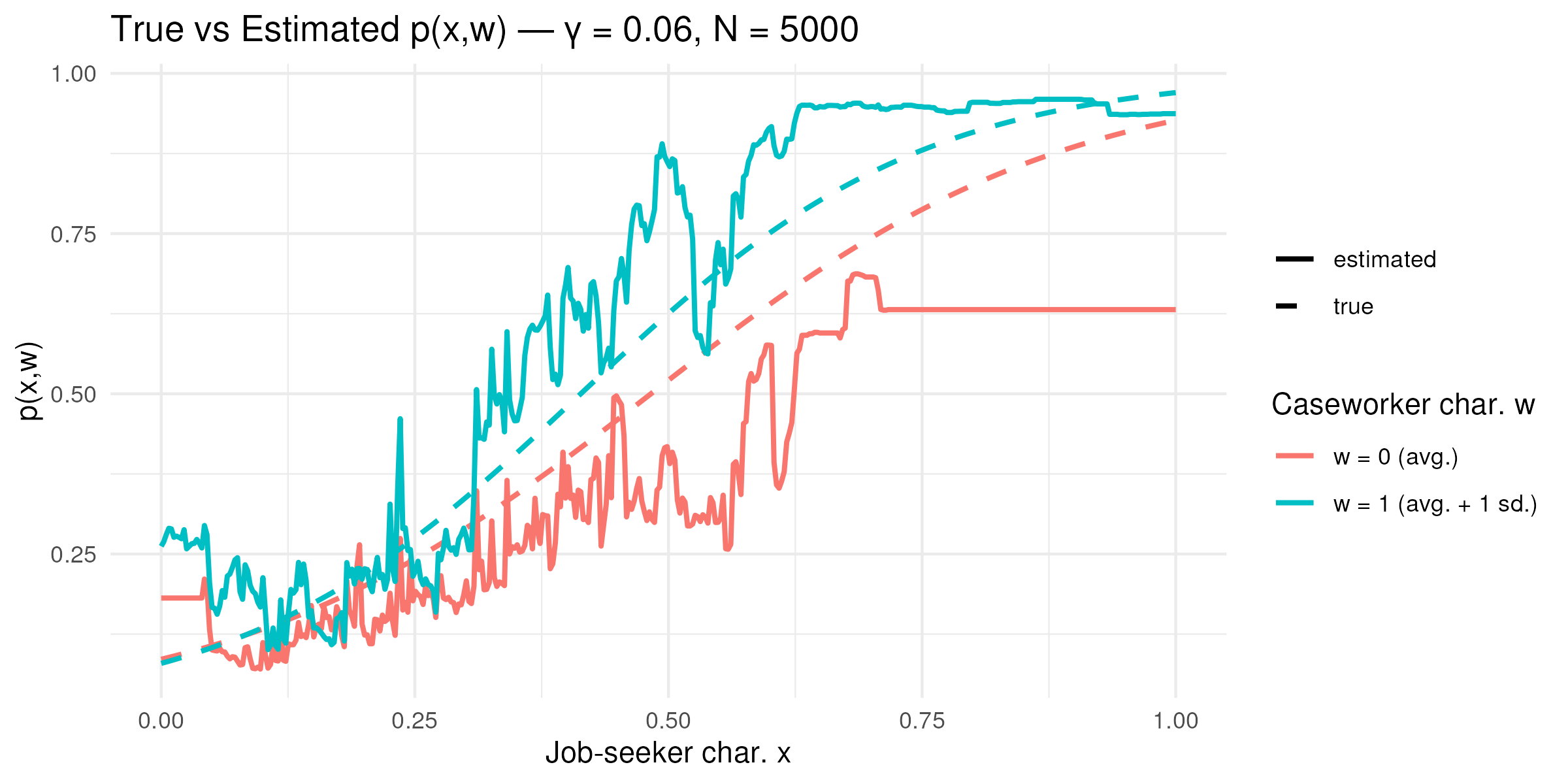}
        \caption{$\hat{c}(x,w)$, $N = 5,000$}
        \label{fig:numex2_hat_c_curve_5000}
    \end{subfigure}
    \begin{subfigure}[t]{0.49\textwidth}
        \centering
        \includegraphics[width=\textwidth]{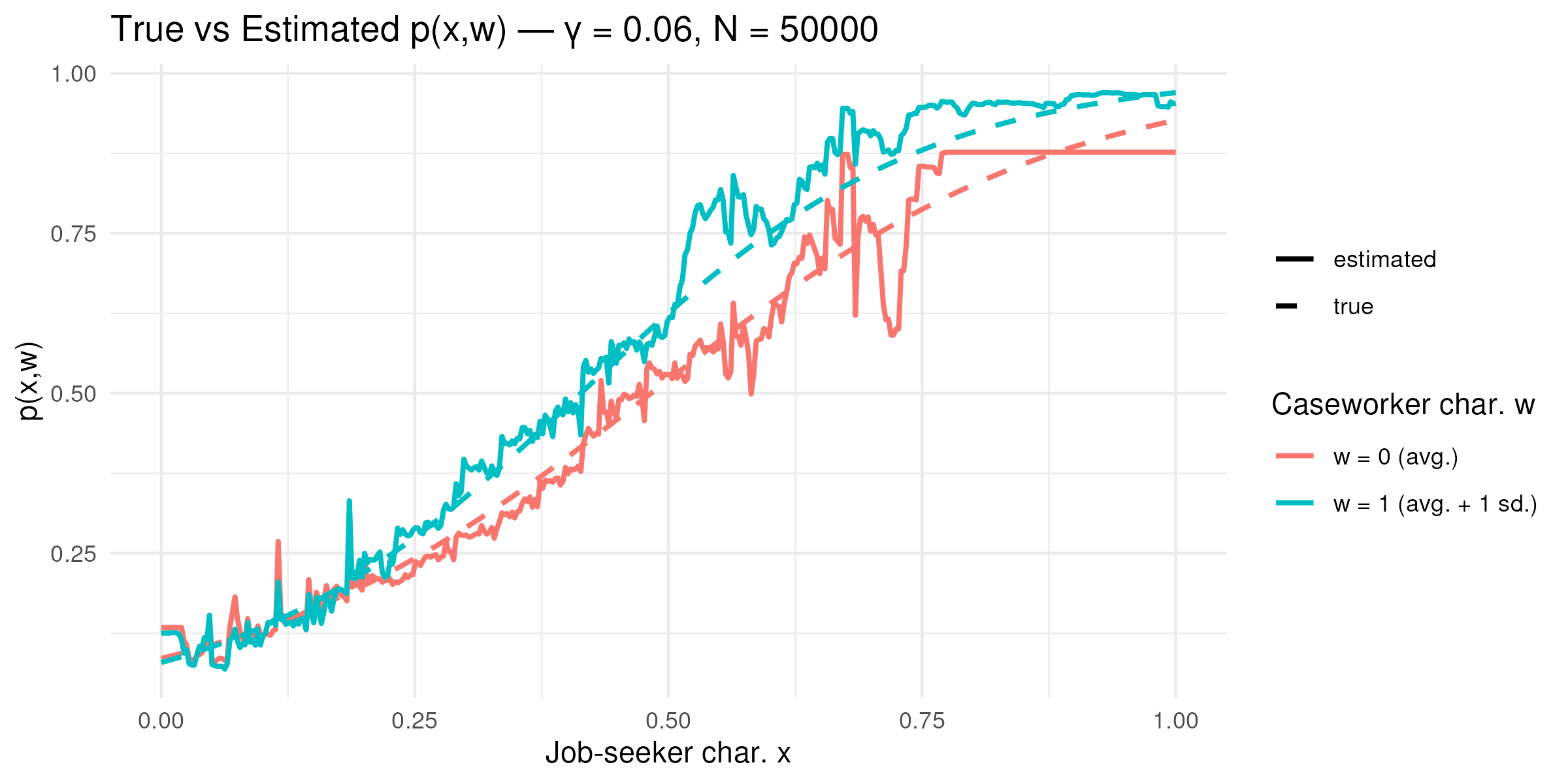}
        \caption{$\hat{c}(x,w)$, $N = 50,000$}
        \label{fig:numex2_hat_c_curve_50000}
    \end{subfigure}
    \vspace{0.5cm}

    
    \begin{subfigure}[t]{0.49\textwidth}
        \centering
        \includegraphics[width=\textwidth]{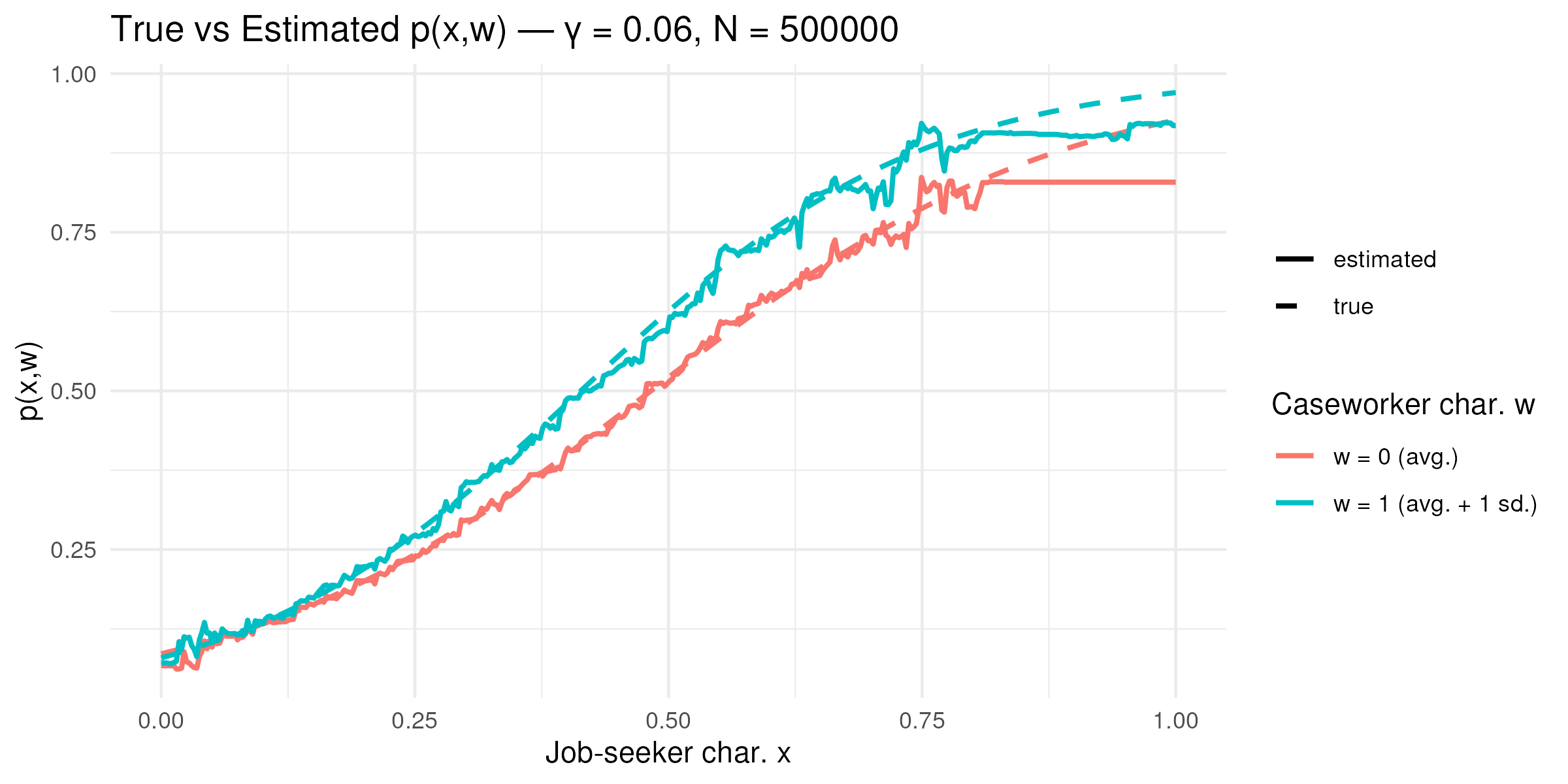}
        \caption{$\hat{c}(x,w)$, $N = 500,000$}
        \label{fig:numex2_hat_c_500000}
    \end{subfigure}
    \begin{subfigure}[t]{0.49\textwidth}
        \centering
        \includegraphics[width=\textwidth]{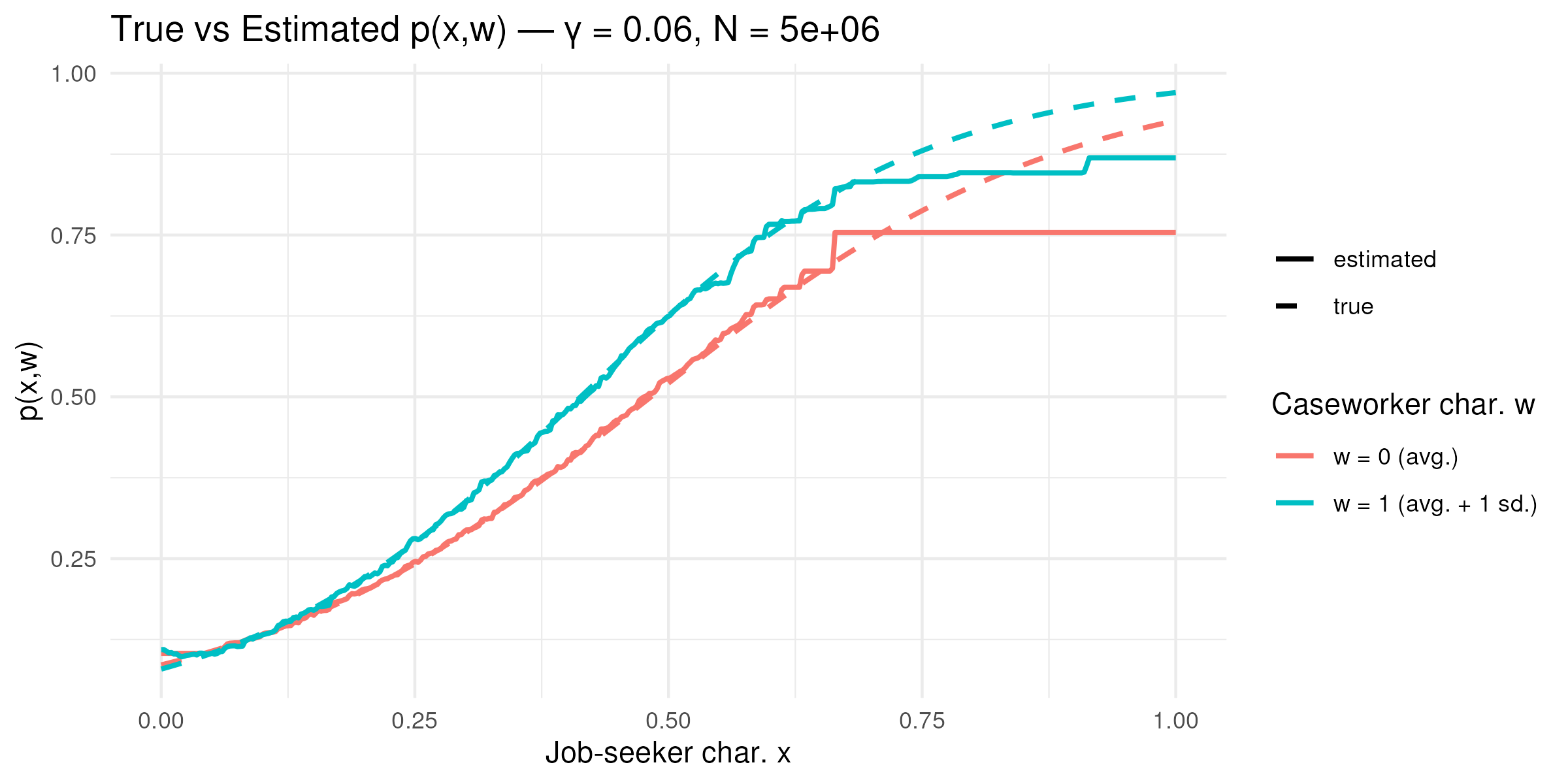}
        \caption{$\hat{c}(x,w)$, $N = 5,000,000$}
        \label{fig:numex2_hat_c_curve_5000000}
    \end{subfigure}

    \caption{Estimated cost functions $\hat{c}(x,w)$ for different training sample sizes, and the true cost function $c(x,w)$ ($\gamma = 0.06$)}
    \label{fig:numex2_cost_curve_estimations}
\end{figure}

To obtain an estimated cost function $\hat{c}(x, w)$, we train an XGBoost model on the training sample, regressing $Y_\ell$ on $(X_\ell, W_\ell, X_\ell \times W_\ell)$. Figures \ref{fig:numex2_cost_estimations} and \ref{fig:numex2_cost_curve_estimations} allow us to compare the true and estimated cost functions as the training sample size $N$ varies. In these figures the difference in the marginal effect of $w$ is held fixed at $\gamma = 0.06$.

\paragraph{Welfare evaluation of the different transport plans}

We compare the average (over training samples) welfare generated by the following matching policies:

\begin{itemize}
    \item $\hat{\pi}^{\mathrm{ROT}}(c)$: welfare attained by a feasible ROT plan---i.e. a plan learned using an estimated cost function $\hat{c}(x, w)$. This depends on the training sample size $N$ and the regularization parameter $1/\eta \in \{0, 0.002, 0.01, 0.05\}$, where $1/\eta = 0$ corresponds to the unregularized OT plan.
    
    \item ${\pi}^{\mathrm{ROT}}(c)$: welfare attained by the oracle ROT plan---i.e. a plan learned based on the (unknown) true cost function $c(x, w)$. This depends on the regularization parameter $1/\eta \in \{0, 0.002, 0.01, 0.05\}$.
    
    \item $(\mu_n \otimes \nu_n)(c)$: welfare attained under random matching, the default policy in this context.
\end{itemize}

The ROT plans are computed using the Sinkhorn algorithm in the log domain for numerical stability. The unregularized OT plan (computed with $1/\eta = 0$) is obtained using the Hungarian assignment algorithm. Simulations are conducted for different values of the training sample size $N$, regularization parameter $1/\eta$, and ``complementarity'' parameter $\gamma$.

\medskip

We report the welfare gain (or loss) of each ROT plan in comparison to random matching (the default policy in the current French Public Employment Services system) in both (i) absolute terms (the pp. change in the job finding rate) and (ii) relative terms (the difference in welfare scaled by the welfare gap between random matching and the oracle unregularized OT plan).
\[
\mu_n \otimes \nu_n)(c) - \hat\pi_n^{\mathrm{ROT}}(c) \quad \text{and} \quad \frac{(\mu_n \otimes \nu_n)(c) - \hat{\pi}_n^{\mathrm{ROT}}(c)}{(\mu_n \otimes \nu_n)(c) - \pi_n^*(c)}.
\]
To approximate the expectation over sampling uncertainty with respect to the training sample, we report the average values of the above quantities over 200 repetitions of the (i)-training set sampling/(ii)-cost estimation/(iii)-transport plan computation procedure.

\medskip

\begin{figure}[H]
\centering
\begin{subfigure}[t]{\textwidth}
        \centering
        \includegraphics[width=\textwidth]{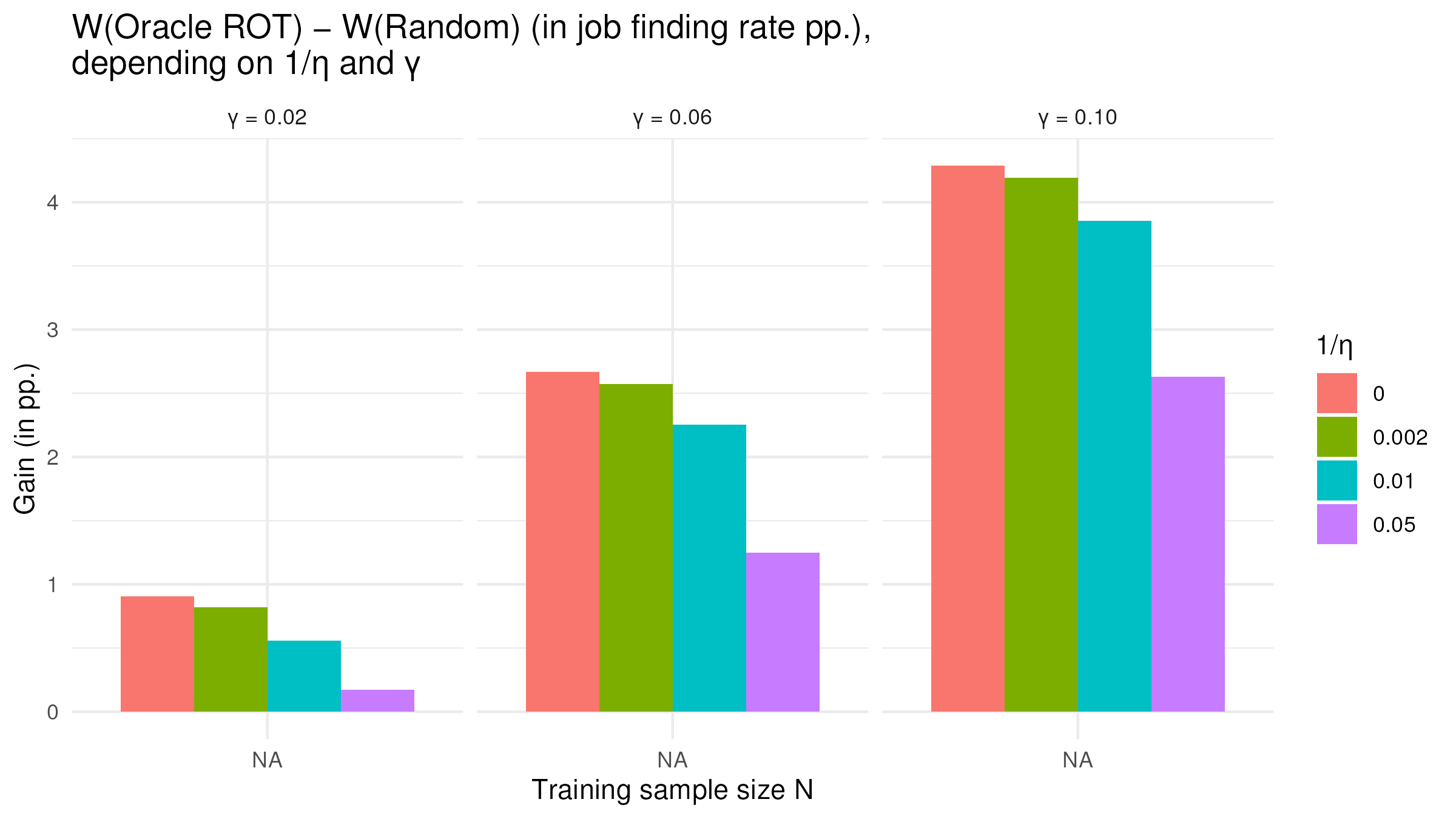}
        \caption{Absolute gain (pp.)}
        \label{fig:numex2_welfare_oracle_rot_pp}
\end{subfigure}
\begin{subfigure}[t]{\textwidth}
        \centering
        \includegraphics[width=\textwidth]{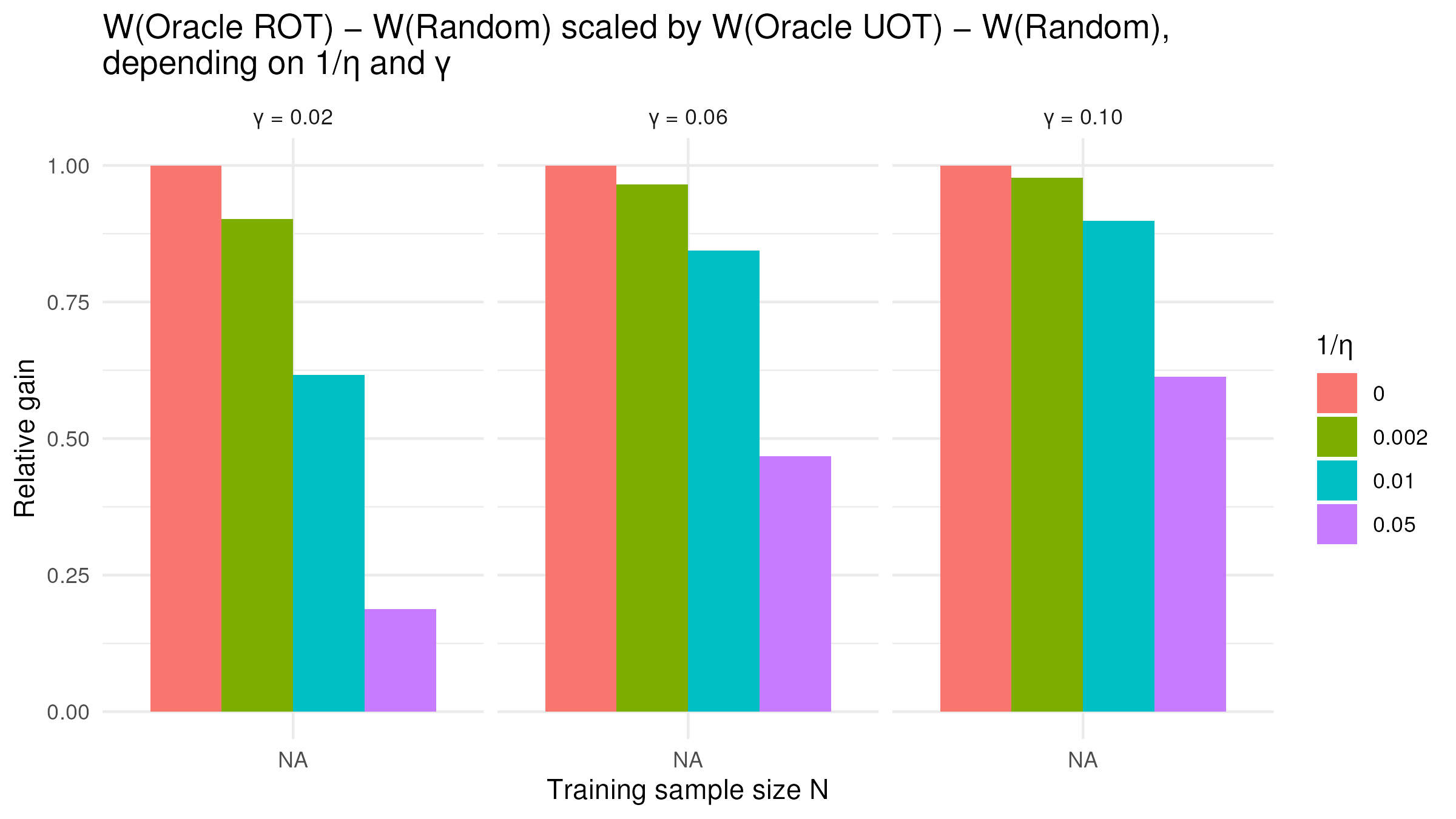}
        \caption{Gain relative to oracle}
        \label{fig:numex2_welfare_oracle_rot_relative}
\end{subfigure}
\caption{Absolute and relative gains from oracle ROT plans}\label{fig:numex2_welfare_oracle_rot}
\caption*{\footnotesize \textit{Notes:} these figures present the absolute (in job finding rate pp.) and relative welfare gain of oracle policies---unregularized and regularized OT plans learned using the (unknown) true cost function $c(x,w)$---compared to the default random matching policy. The relative gain is with respect to the maximal gain generated by the oracle unregularized OT plan. Colors indicate the amount of regularization. The $x$-axis indicates the training sample size---irrelevant here since we consider oracle policies. Panels divide the graphs depending on the complementarity parameter $\gamma$ used to define $c(x,w)$.}
\end{figure}

Before considering feasible/learnable policies, we first discuss (as in the previous numerical example) the performance of the oracle policies---i.e., the unregularized and regularized OT plans learned using the (unknown) true cost function $c(x,w)$,  
\[
\mu_n \otimes \nu_n)(c) - \pi_n^{\mathrm{ROT}}(c) \quad \text{and} \quad \frac{(\mu_n \otimes \nu_n)(c) - \pi_n^{\mathrm{ROT}}(c)}{(\mu_n \otimes \nu_n)(c) - \pi_n^*(c)}.
\]
These are presented in Figure \ref{fig:numex2_welfare_oracle_rot}. An interesting pattern in subfigure \ref{fig:numex2_welfare_oracle_rot_relative} is that the cost of regularization is higher (in relative terms) when complementarities, as ``proxied'' by the parameter $\gamma$, are stronger and the gains from reallocation are higher. In addition, the results shown in subfigure \ref{fig:numex2_welfare_oracle_rot_pp} indicate that even with $\gamma$ as low as $0.02$, the potential gains from reallocation (measured by the absolute gain associated with the unregularized OT plan, $1 / \eta=0$) approach $1$ percentage point. Given the absence of any reallocation cost, this is a sizeable economic gain. In comparison, \cite{behaghel2014} report intention-to-treat estimates for intensive job search counseling programs in France of between 1.5 pp. (se.: 0.8) for private providers and 2.8 pp. (se.: 1.2) for public providers,  and the sizeable costs per program participant lead the authors to adopt a cautious stance on the cost-effectiveness of these programs---at least when privately provided.

\medskip

\begin{figure}[H]
    \centering

    
    \begin{subfigure}[t]{0.49\textwidth}
        \centering 
        \includegraphics[width=\textwidth]{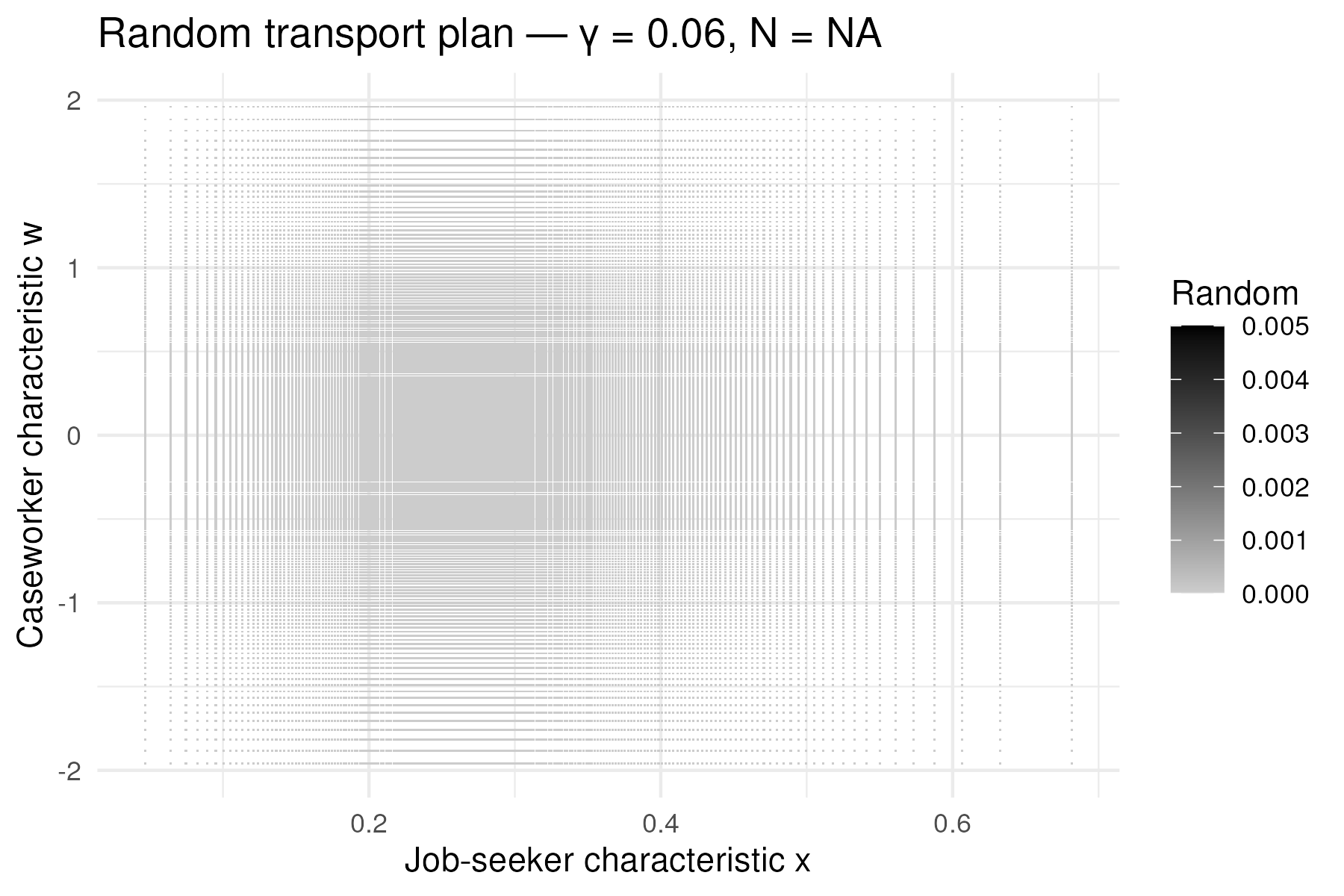}
        \caption{Random matching (for reference)}
    \end{subfigure}
    \begin{subfigure}[t]{0.49\textwidth}
        \centering
        \includegraphics[width=\textwidth]{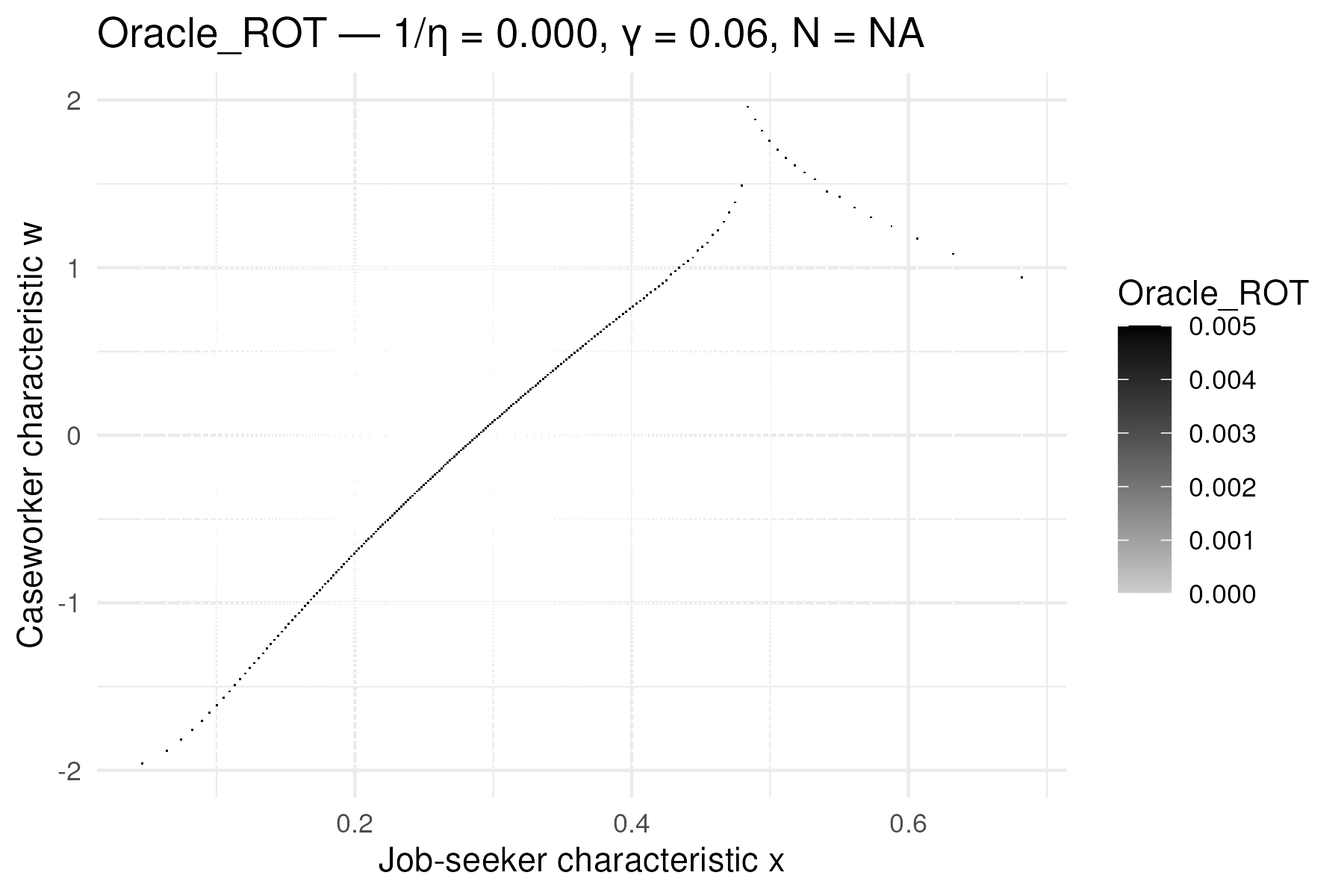}
        \caption{Oracle Unregularized OT}
    \end{subfigure}
    \vspace{0.5cm}

    \begin{subfigure}[t]{0.49\textwidth}
        \centering
        \includegraphics[width=\textwidth]{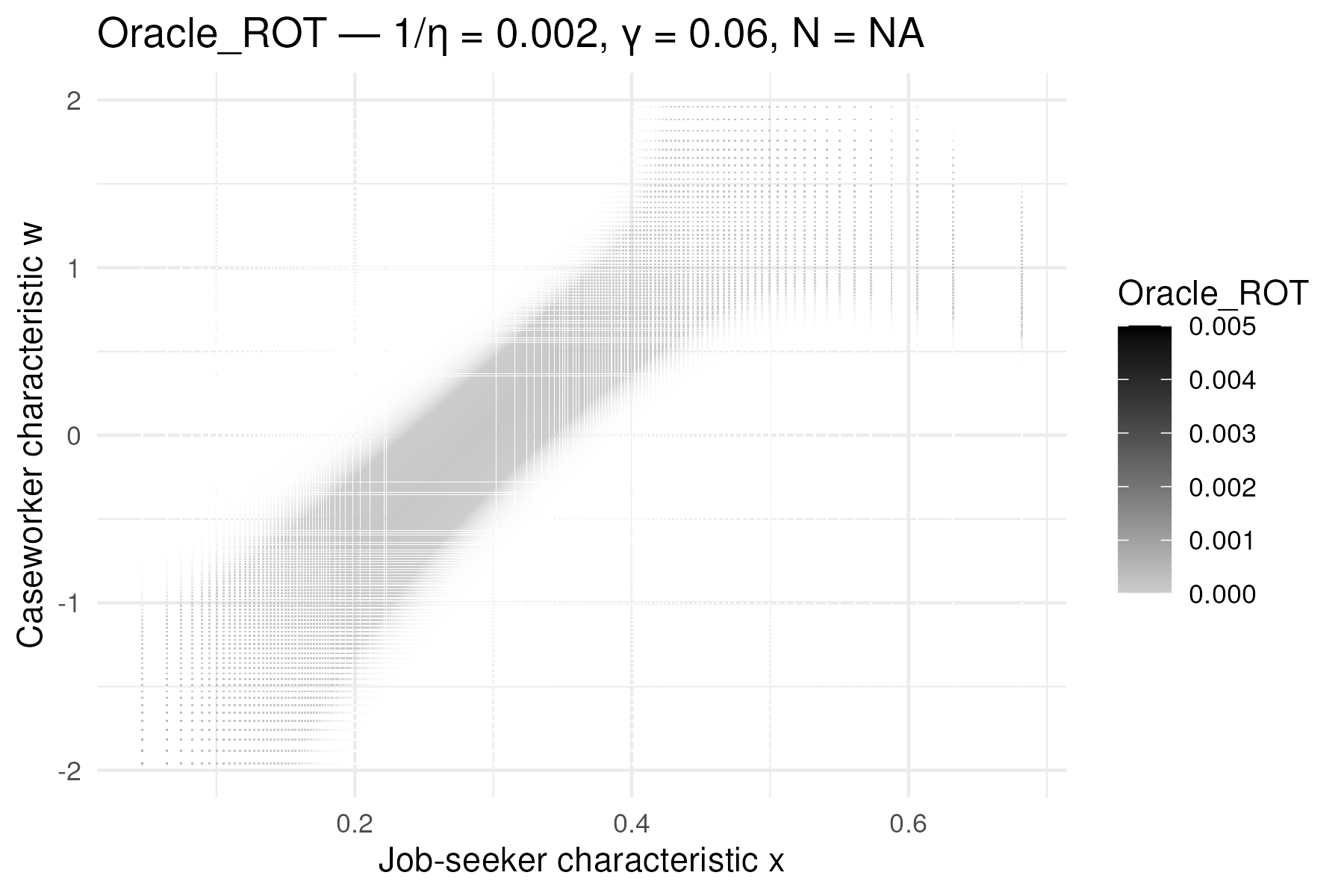}
        \caption{Oracle ROT ($1/\eta = 0.002$)}
    \end{subfigure}
    \begin{subfigure}[t]{0.49\textwidth}
        \centering
        \includegraphics[width=\textwidth]{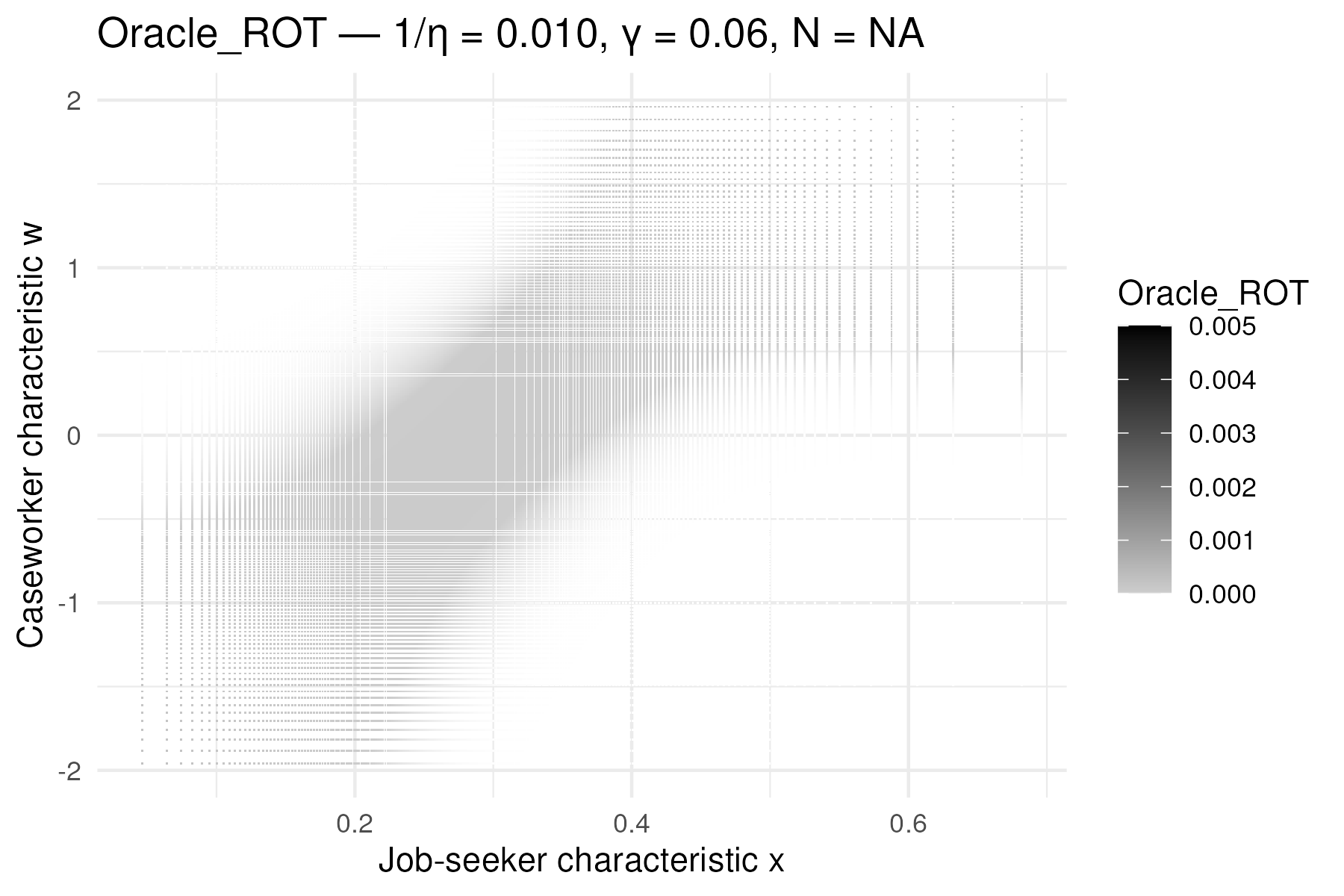}
        \caption{Oracle ROT ($1/\eta = 0.010$)}
    \end{subfigure}

    \caption{Expected (over training set samples) joint distribution of $(X,W)$ implied by the random and select oracle transport plans, with complementarities set at $\gamma=0.06$}
    \label{fig:numex2_hm_oracle_plans}
\end{figure}

Figure \ref{fig:numex2_hm_oracle_plans} presents heatmaps describing the joint distribution of $(X,W)$ implied by the oracle transport plans. As expected given the complementarities featured in the true cost function, these oracle plans place most probability mass on the diagonal of the product space. This is optimal as the marginal effect of increasing the caseworker characteristic $w$ on the job finding probability is increasing with the jobs seekers' characteristic $x$ over most of its support, ---i.e., $\frac{\partial^2}{\partial x\partial w} p(x,w) > 0$ is increasing in $x$ over most of the support of $X$. However, the oracle unregularized OT plan is not pure positive assortative matching---as can be seen in panel (b) of Figure \ref{fig:numex2_hm_oracle_plans}. This is because we set the job finding probability function to be a logistic function.\footnote{Job seekers with $x \approx 0.5$ have a job finding probability equal to $0.5$ when matched with a caseworker whose characteristic is equal to the mean of the $W$ distribution (i.e. $w=0$). Since the derivative of the logistic function with respect to its index is highest at index $= \operatorname{logistic}^{-1}(0.5)$, in this setting the unregularized optimal plan assigns the top caseworkers to job seekers with $x \approx 0.5$.}

\medskip

\begin{figure}[H]
    \centering

    
    \begin{subfigure}[t]{0.49\textwidth}
        \centering 
        \includegraphics[width=\textwidth]{figures/plan_heat_Random_gap06pp_NNA.png}
        \caption{Random matching (for reference)}
    \end{subfigure}
    \begin{subfigure}[t]{0.49\textwidth}
        \centering
        \includegraphics[width=\textwidth]{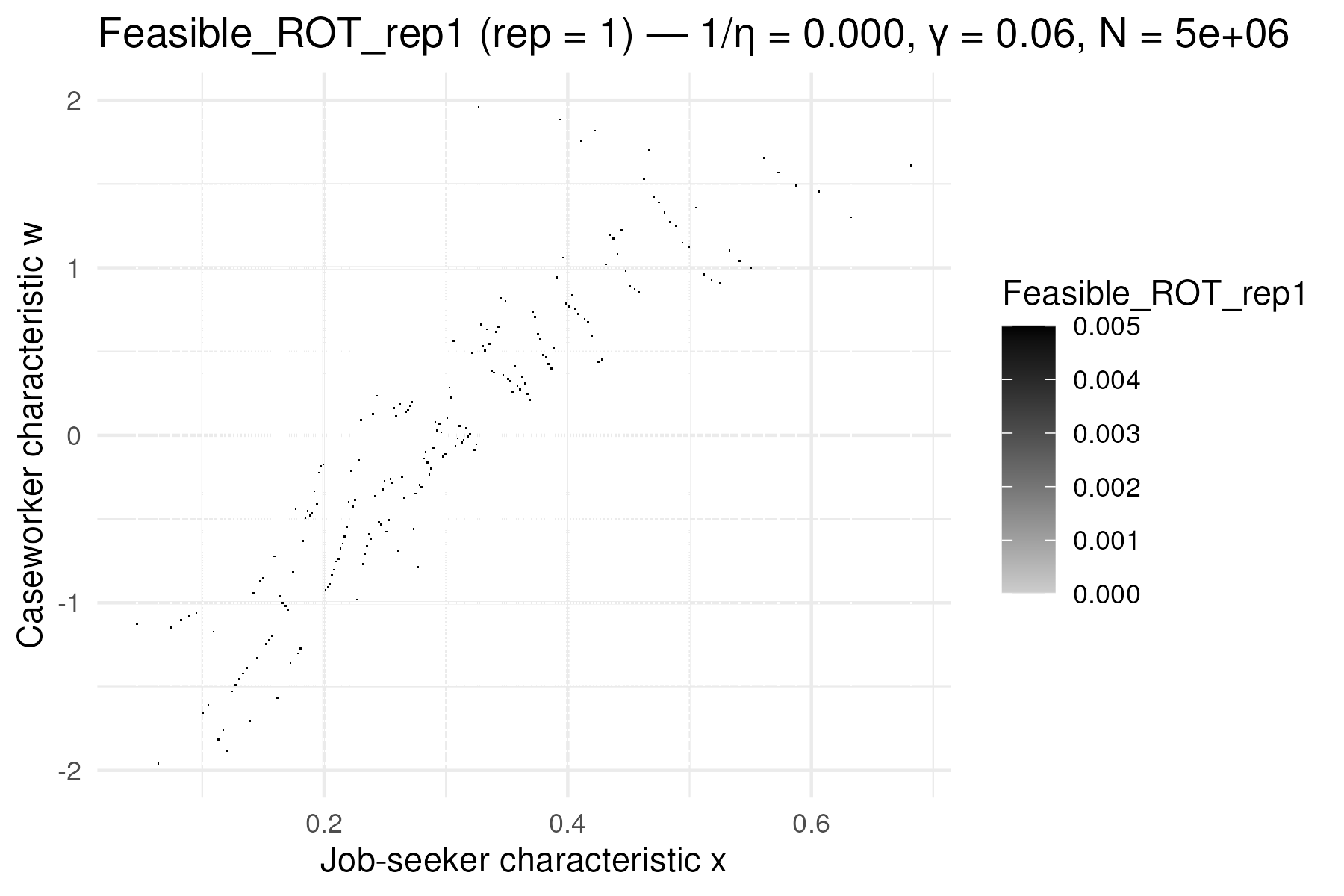}
        \caption{Feasible Unregularized OT}
    \end{subfigure}
    \vspace{0.5cm}

    \begin{subfigure}[t]{0.49\textwidth}
        \centering
        \includegraphics[width=\textwidth]{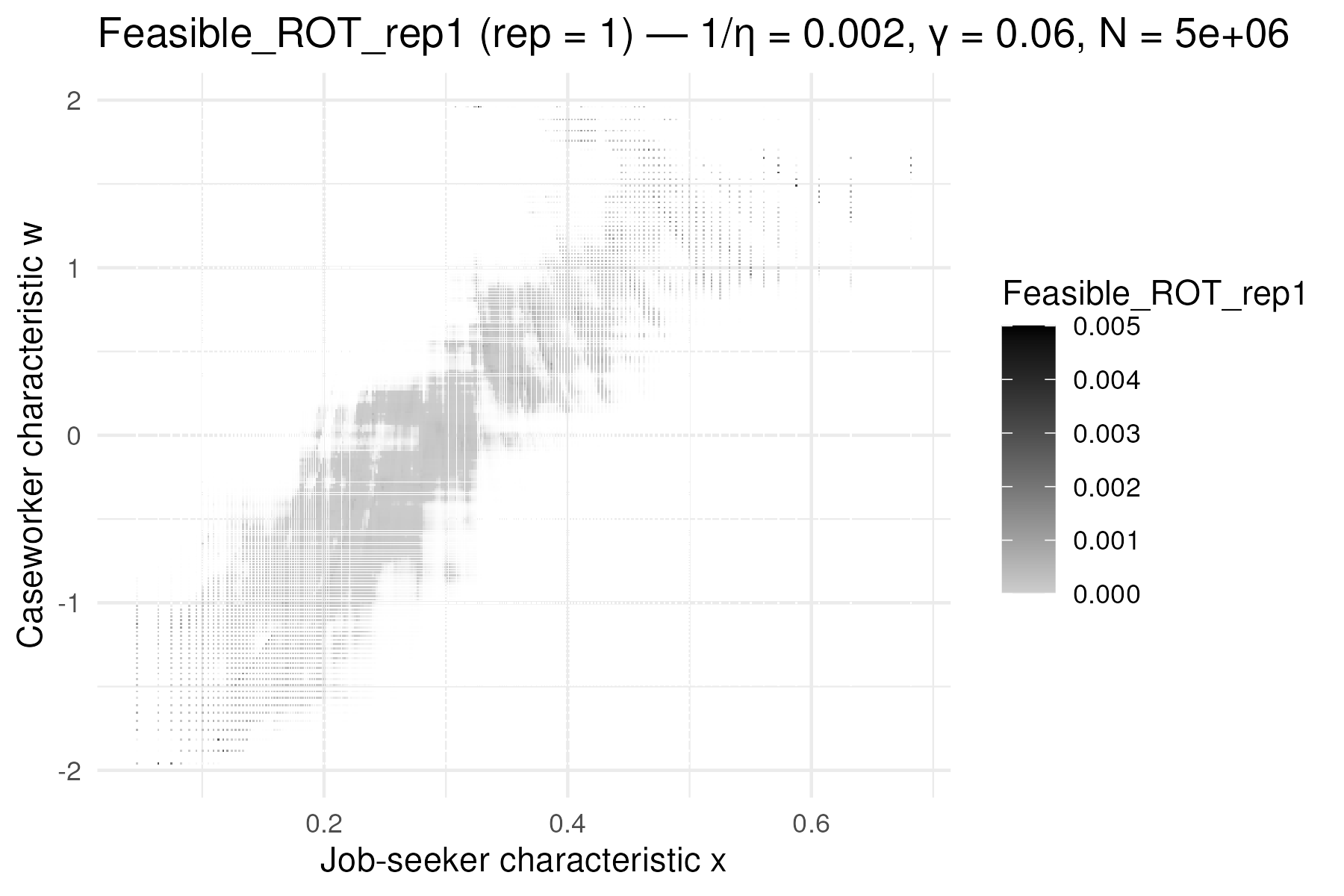}
        \caption{Feasible ROT ($1/\eta = 0.002$)}
    \end{subfigure}
    \begin{subfigure}[t]{0.49\textwidth}
        \centering
        \includegraphics[width=\textwidth]{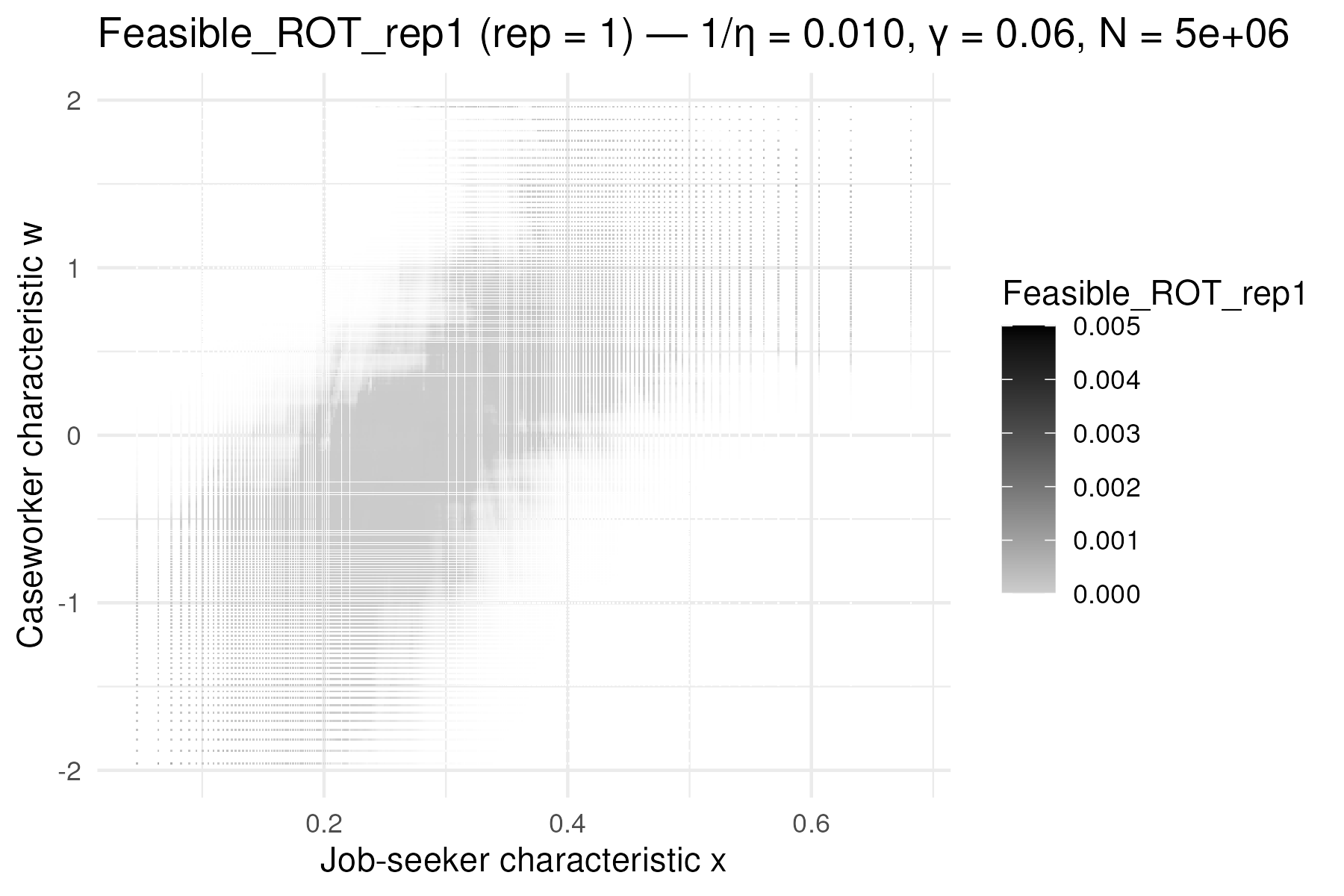}
        \caption{Feasible ROT ($1/\eta = 0.010$)}
    \end{subfigure}

    \caption{Joint distribution of $(X,W)$ implied by the random vs. various feasible transport plans, for complementarities set at $\gamma=0.06$ (cost function estimated with $N=5,000,000$)}
    \label{fig:numex2_hm_feasible_plans}
\end{figure}

Now, we can turn to the performance of policies learned using estimated cost functions. Figure \ref{fig:numex2_hm_feasible_plans} presents similar graphs to Figure \ref{fig:numex2_hm_oracle_plans} for these feasible policies for one iteration of the (i)-training set sampling/(ii)-cost estimation/(iii)-transport plan computation procedure---with the cost function estimated using a training sample of $5,000,000$ observations. Reassuringly, when the cost function is estimated from a training dataset of this size, the resulting plans are quite similar to their oracle counterparts in Figure \ref{fig:numex2_hm_oracle_plans}. That said, the unregularized OT plan shown in panel (b) is noisier than its oracle counterpart, and appears to place some mass on a comparable support to the regularized plan shown in panel (c).

\begin{figure}[H]
\centering
\begin{subfigure}[t]{\textwidth}
        \centering
        \includegraphics[width=\textwidth]{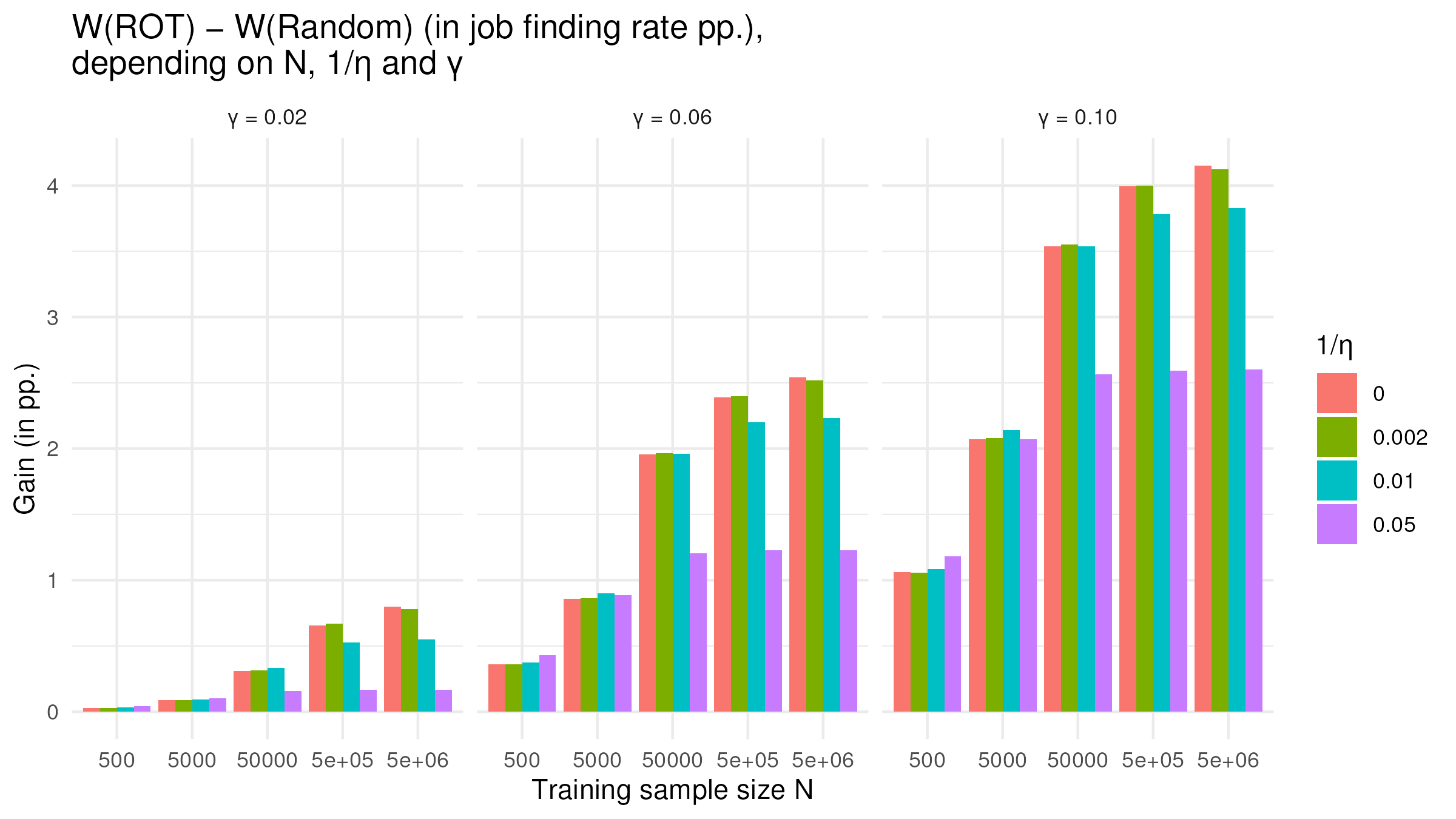}
        \caption{Gain in pp.}
        \label{fig:numex2_welfare_est_rot_pp}
\end{subfigure}
\begin{subfigure}[t]{\textwidth}
        \centering
        \includegraphics[width=\textwidth]{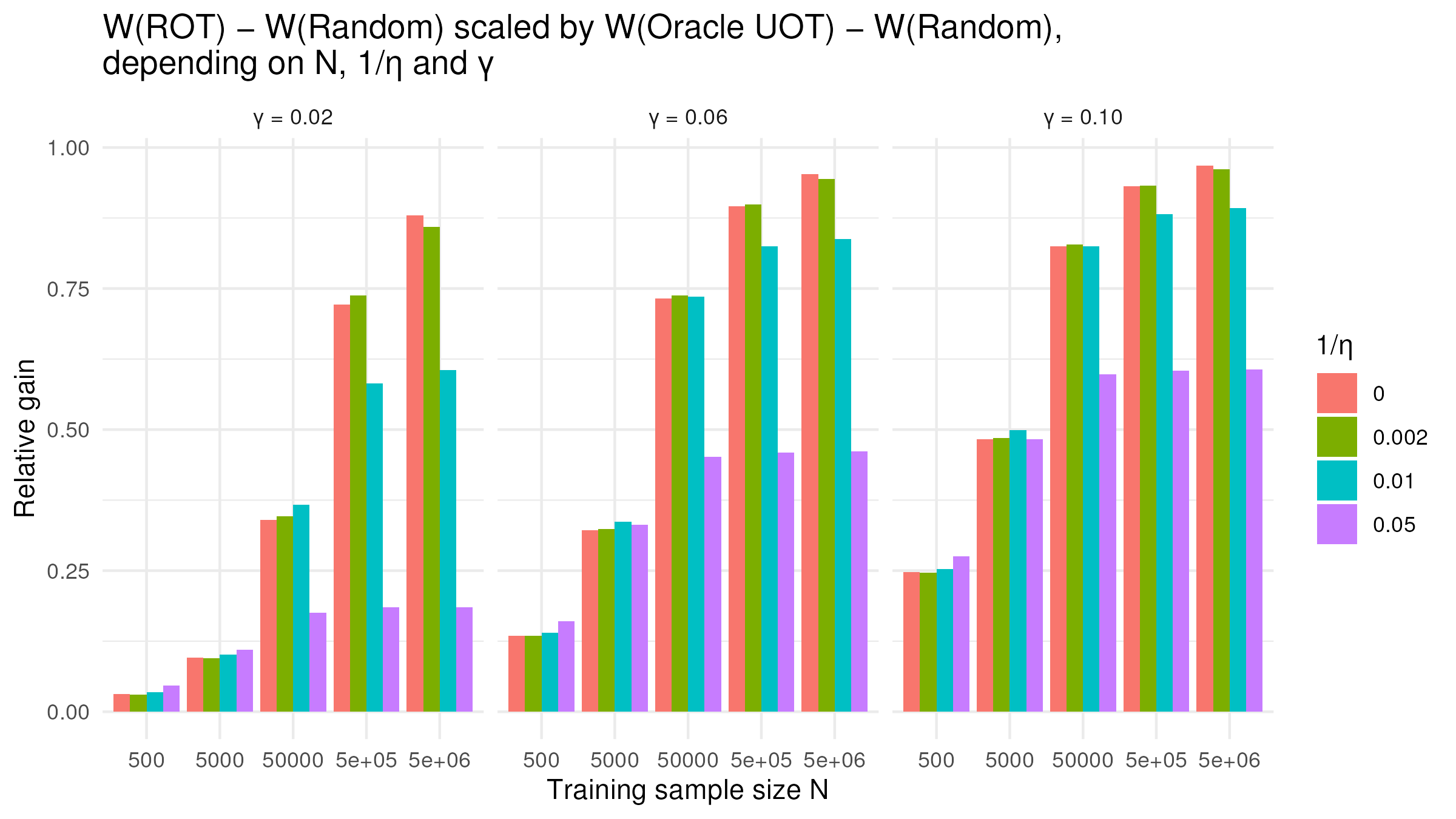}
        \caption{Gain relative to oracle}
        \label{fig:numex2_welfare_est_rot_relative}
\end{subfigure}
\caption{Absolute and relative gains from oracle ROT plans}\label{fig:numex2_welfare_est_rot}
\caption*{\footnotesize \textit{Notes:} these figures present the absolute (in job finding rate pp.) and relative welfare gain of feasible policies---unregularized and regularized OT plans learned using the estimated cost function $\hat c(x,w)$---compared to the default random matching policy. The relative gain is with respect to the maximal gain generated by the oracle unregularized OT plan. Colors indicate the amount of regularization. The $x$-axis indicates the training sample size used to estimate the cost function. Panels divide the graphs depending on the complementarity parameter $\gamma$ used to define $c(x,w)$.}
\end{figure}

Figure \ref{fig:numex2_welfare_est_rot} reports similar quantities to Figure \ref{fig:numex2_welfare_oracle_rot} for feasible (R)OT plans learned using an estimated cost function. Overall, Figure \ref{fig:numex2_welfare_est_rot} parallels Figure \ref{fig:numex2_welfare_oracle_rot}. Unlike the first numerical example, there is no clear pattern in terms of the benefits (or costs) of regularization for smaller training samples.\footnote{Although some amount of regularization appears to be, if anything, costless for the smallest training sample sizes.} For larger training sample sizes, the welfare gain from learned policies approaches that of the oracle unregularized optimal plan.

\section{Conclusion}

This paper studies policy learning for two-sided matching policies. We propose the use of entropy regularized empirical optimal transport and assess its welfare performance when matching caseworkers and job seekers in the context of a data generating processes calibrated to match French administrative data. 

We leave several questions for future research. First, the performance of the estimated regularized policy can be sensitive to the choice of penalty parameter, and there is currently no method to tune it in a data-driven manner. Second, in our welfare regret analysis we obtain only an upper bound. A lower bound for regret remains to be investigated. Third, we set the base measure of the entropy penalty to independent coupling. Depending on the economic interpretation attached to the penalty term, it may be sensible to choose a different base.  Investigation of the economic justification for the choice of penalty, in terms of both the base and the divergence measure, is left for future work. Future versions of this paper will also attempt to apply the method to real-world administrative data on the matching of job seekers and caseworkers to go beyond the simulations currently presented.    

\section*{Appendix}

\appendix

\section{Proofs}\label{app:proofs}

\subsection{Lemma \ref{lemma:bounds_for_pfg}}
\begin{proof}
Given the definition of $\hat{p}_n(x,w)$, the constraint that the marginal distribution of $X$ must coincide with the empirical distribution at each $x_i$ can be written as $\nu_n(\hat{p}_n(x_i,w)) = \frac{1}{n}$. Using the definition of $\hat{p}_n(x,w)$ in the constraint and simplifying gives
\begin{align*} 
1 & = \frac{1}{n} \sum_{j=1}^n \exp \left( - \eta \hat{c}(x_i, w_j) + \eta \hat{f}_n(x_i) + \eta \hat{g}_n(w_j) \right) \\
& \geq \exp \left( - \eta \bar{c} + \eta \hat{f}_n(x_i) \right) \cdot \frac{1}{n} \sum_{j=1}^n  \exp (\eta \hat{g}_n(w_j)) \\
& \geq \exp(- \eta \bar{c} + \eta \hat{f}_n(x_i)),
\end{align*}
where the first inequality follows by Assumption 1, and the second inequality follows by Jensen's inequality and the location normalization (\ref{eq:location normalization g}) for $\hat{g}_n$. As a result, we obtain 
\begin{equation} \label{eq:fn upper bound}
\hat{f}_n (x_i) \leq \bar{c}
\end{equation}
for all $i \in \mathcal{I}$. Similarly, the constraint that the marginal distribution over $W$ must coincide with the empirical distribution at each $w_j$ can be written as $\mu_n(\hat{p}_n(x,w_j)) = \frac{1}{n}$, and from this we can obtain $\hat{g}_n (w_j) \leq \bar{c}$ for all $j \in \mathcal{J}$.

Next, to obtain a lower bound for $\hat{g}_n(w_j)$, consider the constraint $\mu_n(\hat{p}_n(x,w_j)) = \frac{1}{n}$, 
\begin{align*} 
1 & = \frac{1}{n} \sum_{i=1}^n \exp \left( - \eta \hat{c}(x_i, w_j) + \eta \hat{f}_n(x_i) + \eta \hat{g}_n(w_j) \right) \\
& \leq \exp \left( \eta \bar{c} + \eta \hat{g}_n(w_j) \right),
\end{align*}
where the inequality follows by Assumption 1 and (\ref{eq:fn upper bound}). Hence, we obtain
\begin{equation} \label{eq:gn lower bound}
\hat{g}_n(w_j) \geq  - \bar{c},
\end{equation}
for all $j \in \mathcal{J}$. We repeat the same argument to obtain $\hat{f}_n(x_i) \geq  - \bar{c}$ for all $i \in \mathcal{I}$. 

Plug the uniform upper and lower bounds for $\hat{f}_n(x_j)$ and $\hat{g}_n(w_j)$ and the bounds for $\hat{c}(x_i,w_j)$ into the definition of $\hat{p}_n(x_i,w_j)$ to obtain the bounds for the density of the regularized optimal coupling, 
\begin{equation*}
\exp(-3\eta \bar{c}) \leq \hat{p}_n(x_i,w_j) \leq \exp(2 \eta \bar{c}).
\end{equation*}
\end{proof}

\subsection{Proposition \ref{prop:regret_ROT}}
\begin{proof}
Since $\hat{\pi}_n^{ROT}(x_i,y_j) = \frac{1}{n^2} \exp(-\eta \hat{c}(x_i,y_j) + \eta \hat{f}_n(x_i) + \eta \hat{g}_n(y_j))$, we have
\begin{equation}
| \hat{\pi}_n^{ROT}(c(x,y)) - \hat{\pi}_n^{ROT}(\hat{c}(x,y)) | = (\mu_n \otimes \nu_n)(|\hat{c} - c| \hat{p}_n) \leq e^{2 \eta \bar{c}} (\mu_n \otimes \nu_n)(|\hat{c} - c|),
\end{equation}
where the inequality follows by Lemma \ref{lemma:bounds_for_pfg}. This proves (\ref{eq:prop1_bound1}).

To show (\ref{eq:prop1_bound2}), take a first-order mean value expansion of $\Phi_n((1-t)c+t\hat{c}, \hat{f}_n, \hat{g}_n)$ in $t$. Since 
\begin{equation}
\frac{d}{dt} \Phi_n((1-t)c+t\hat{c}, \hat{f}_n, \hat{g}_n) = (\mu_n \otimes \nu_n) \left( (\hat{c} - c) \exp \left\{ - \eta[ (1-t)c + t \hat{c}] + \eta f + \eta g  \right\} \right),
\end{equation}
for $\tilde{t} \in [0,1]$, we have
\begin{align}
|\Phi_n(\hat{c}, \hat{f}_n, \hat{g}_n) - \Phi_n(c, \hat{f}_n, \hat{g}_n) | & =  (\mu_n \otimes \nu_n) \left( (\hat{c} - c)^2 \exp \left\{ - \eta[ (1-\tilde{t})c + \tilde{t} \hat{c}] + \eta \hat{f}_n + \eta \hat{g}_n  \right\} \right) \\
& \leq e^{2 \eta \bar{c}} \| \hat{c} - c \|^2_{L^2(\mu_n \otimes \nu_n)},
\end{align}
where the inequality follows by Lemma \ref{lemma:bounds_for_pfg}. This proves (\ref{eq:prop1_bound2}).

The other bounds (\ref{eq:prop1_bound3}) - (\ref{eq:prop1_bound5}) follow by noting that $E_P^n[ (\mu_n \otimes \nu_n) (\cdot) ] = (\mu \otimes \nu)(\cdot)$.

\end{proof}

\subsection{Proposition \ref{prop:regularization_bias}}
\begin{proof}
Since $\pi_n^{ROT}$ minimizes the regularized average cost, replacing $\pi_n^{ROT}$ with $\pi_n^{\ast}$ in the regularization bias gives an upper bound;
\begin{equation}
\pi_n^{ROT}(c(x,y)) + \frac{1}{\eta} KL( \pi_n^{ROT} \| \mu_n \otimes \nu_n ) - \pi_n^{\ast}(c(x,y)) \leq \frac{1}{\eta} KL( \pi_n^{\ast} \| \mu_n \otimes \nu_n ).
\end{equation}
Since $\pi_n{\ast}$ is the solution of a linear program, and the extreme points of the constraint set is a polyhedron with extremeum points corresponding to degenerate permutation distributions, $\pi_n^{\ast}$ is supported only on $n$ points with equal probability masses $1/n$. Hence, noting that $\lim_{a \to 0} a \cdot \log a = 0$, we obtain
\begin{equation}
 \frac{1}{\eta} KL( \pi_n^{\ast} \| \mu_n \otimes \nu_n ) = \frac{1}{\eta}\sum_{i=1}^n \sum_{j=1}^n \pi_n^{\ast}(x_i,y_j) \log \left( \frac{\pi_n^{\ast}(x_i,y_j)}{1/n^2} \right) = \frac{1}{\eta n} \sum_{ \{(i,j): \pi_n^{\ast}(x_i,y_j) > 0 \} } \log n = \frac{\log n}{ \eta }. \notag
\end{equation}
\end{proof}

\bibliographystyle{ecta}
\bibliography{biblio.bib}

\end{document}